\documentclass[letterpaper, 12pt, draftcls, onecolumn]{ieeeconf}
\IEEEoverridecommandlockouts                              % This command is only needed if 
\usepackage[dvipdfmx]{graphicx}% for pdf, bitmapped graphics files
\usepackage{amsmath, amssymb, bm, mathrsfs}
\usepackage{cite, color, graphicx, subcaption}

\newcommand{\Int}{\mathop{\rm Int~\!}}
\newcommand{\Cl}{\mathop{\rm Cl~\!}}

\newtheorem{thm}{Theorem}[section]
\newtheorem{prop}[thm]{Proposition}
\newtheorem{defn}[thm]{Definition}

\newtheorem{assum}[thm]{Assumption}

\newtheorem{lem}[thm]{Lemma}
\newtheorem{rem}[thm]{Remark}
\newtheorem{cor}[thm]{Corollary}
\newtheorem{alg}[thm]{Algorithm}

\title{\LARGE \bf
Stability Analysis of Sampled-Data Switched Systems \\ 
with Quantization
}

\author{Masashi Wakaiki and Yutaka Yamamoto
	\thanks{
This work was supported in part by The Kyoto University Foundation.
The material in this paper was partially presented at the
19th IFAC World Congress, August 24-29, 2014, South Africa.
	}
	\thanks{
		M. Wakaiki is with the Center for Control,
		Dynamical-systems and Computation (CCDC), University of California,
		Santa Barbara, CA 93106-9560 USA
		(e-mail:{\tt  \ masashiwakaiki@ece.ucsb.edu}).
		Y. Yamamoto is with the Department of Applied Analysis and Complex
		Dynamical Systems, Graduate School of Informatics, Kyoto University, Kyoto
		606-8501, Japan
		(e-mail: {\tt \ yy@i.kyoto-u.ac.jp}).
	}%
}

\begin{document}

	\maketitle
	\thispagestyle{empty}
	\pagestyle{empty}

\begin{abstract}                          % Abstract of not more than 200 words.
We propose a stability analysis method for sampled-data switched linear
systems with finite-level static quantizers.
In the closed-loop system, information on the active mode of the plant
is transmitted to the controller only at each sampling time.
This limitation of switching information leads to a mode mismatch
between the plant and the controller, and 
the system may become unstable.
A mode mismatch also makes it difficult to find an attractor set
to which the state trajectory converges.
A switching condition for stability 
is characterized by
the total time
when the modes of the plant and the controller are different.
Under the condition,
we derive an 
ultimate bound on the state trajectories
by using a common Lyapunov function
computed from a randomized algorithm.
The switching condition can be reduced to
a dwell-time condition.

\end{abstract}

\section{Introduction}
The recent advance of networking technologies makes
control systems more flexible.
%and maintenance/installation-friendly.
However, the use of networks also raises
new challenges such as packet dropouts, variable transmission delays,
and real-time task scheduling. 
Switched system models provide
a mathematical framework for such network properties 
because of their versatility to include
both continuous flows and discrete jumps;
see \cite{Lin2005, Donkers2011 ,Zhang2010, Song2008} and references therein
for the application of switched system models to networked control systems.

On the other hand, many control loops in a 
practical network
contain channels over which only a finite number of bits can be transmitted.
We need to quantize data before sending them out
through a network.
Therefore the effect of data quantization should be taken into consideration to 
achieve stability and desired performance. 
In addition to the practical motivation, 
literature such as \cite{Wong1999,Tatikonda2004, 
	Nair2003Automatica, Okano2014}
has answered the theoretical question of
how much information is necessary/sufficient
for a given control problem.

Switched systems and quantized control have been 
studied extensively but separately;
see, e.g., \cite{Liberzon2003Book, Shorten2007, Lin2009} for switched systems
and \cite{Nair2007,Ishii2012, Matveev2009} for quantized control.
However, quantized control of switched systems has received increasing
attention in recent years.
For discrete-time Markovian jump linear systems,
control problems with limited information have been studied in 
\cite{Nair2003, Liu2009, Ling2010, Xiao2010, Xu2013}.
Also, our previous work \cite{WakaikiMTNS2014} has
investigated the output feedback stabilization of continuous-time
switched systems under a slow-switching assumption.
In most of the above studies, 
the switching behavior of the plant
is available to the controller {\em at all times.}

In contrast, 
in {\it sampled-data} 
switched systems with quantization,
the controller receives 
the quantized measurement and 
the active mode of the plant
{\em only at each sampling time}.
Since 
the controller side does not know
the active mode of the plant
between sampling times,
we do not always use the controller mode 
consistent with the plant mode at
the present time.
The closed-loop system may therefore become unstable when switching occurs between sampling times.
Moreover,
for the stability of quantized systems, 
it is important to obtain regions to which 
the state belongs.
However, mode mismatches yield complicated 
state trajectories, which make it 
difficult to find such regions.

Stabilization of sampled-data switched system
with {\em dynamic} quantizers 
has been first addressed in
\cite{Liberzon2014}, which
has proposed an encoding strategy for state feedback
control. 
This encoding method has been extended to the output feedback case
\cite{Wakaiki2014CDC}
and to the case with disturbances \cite{Yang2015ACC}.
A crucial ingredient in the dynamic quantization 
is a reachable set of the state trajectories 
through sampling intervals.
Propagation of reachable sets is used to
set the quantization values at the 
next sampling time, and the dynamic quantizer
achieves increasingly higher precision as
the state approaches the origin.
On the other hand, we study the stability analysis
of sampled-data switched systems
with {\em finite-level static} quantizers.
For such a closed-loop system,
asymptotic stability cannot be guaranteed.
The objective of the present paper is 
therefore to find an ultimate bound
on the system trajectories as in
the single-modal case, e.g., \cite{Ishii2002Book, Ishii2004, Haimovich2007, Elia2001}.
Since frequent mode mismatches make
the trajectories diverge, a certain 
switching condition is required for
the existence of ultimate bounds. 

As in \cite{Ma2015}
for switched systems with time delays,
we here characterize switching behaviors by
the total time when the controller mode is 
not synchronized with the plant one, which
we call the {\em total mismatch time}.
We
derive a sufficient condition on the total mismatch
time for the system to be stable, by
using an upper bound on the error
due to sampling and quantization.
Moreover, an ultimate bound
on the state trajectories 
is obtained 
under the switching
condition.
For the stability analysis, 
we use a common Lyapunov function
that guarantees stability for all individual modes
in the non-switched case.
We find such Lyapunov functions in
a computationally efficient and less conservative way by
combining the randomized algorithms
in \cite{Ishii2004, Liberzon2004} together.

From the total mismatch time, we can obtain 
an asynchronous switching time ratio.
If the controller mode is synchronized with
the plant one, then
the closed-loop system is stable.
Otherwise, the system may be
unstable.
Hence the total mismatch time is a
characterization similar to
the total activation time ratio \cite{Zhai2001}
between stable modes and unstable ones.
The crucial difference is that 
the unstable modes we consider
are caused by switching within sampling intervals.
Using the dependence of the instability on the sampling period,
we can reduce the switching condition on the total mismatch time
to a dwell-time condition, which is widely used for
the stability analysis of switched systems.
In Section~4, 
we will discuss in detail 
the relationship between the total mismatch time
and the dwell time of switching behaviors.

This paper is organized as follows. 
In Section~2, we present the closed-loop system,
the information structure, and basic assumptions.
In Section~3, we first investigate 
%an upper bound on 
the growth rate of the common
Lyapunov function in the case when switching occurs in a sampling interval.
Next we derive
an ultimate bound on the state,
together with a sufficient condition on switching for stability.
Section~4 is devoted to reduce the derived switching condition to a dwell-time condition. 
We illustrate the results through a numerical example in Section~5.
Finally, concluding remarks are given in Section~6.

This paper is based on a conference paper \cite{Wakaiki2014IFAC}.
In the conference version, some of the proofs 
were omitted due to space limitations.
The present paper provides complete results on the 
stability analysis in addition to an illustrative numerical example.
We also made structural improvements in this paper.

\noindent
{\bf Notation} \\
We denote by $\mathbb{Z}_+$ the set of non-negative integers
$\{k \in \mathbb{Z}:~k \geq 0\}$.
For a set $\Omega \subset \mathbb{R}^{\sf n}$, $\Cl (\Omega)$, 
$\Int (\Omega)$, and $\partial \Omega$ are its closure,
interior, and boundary, respectively.
For sets $\Omega_1, \Omega_2$, 
let $\Omega_1 \setminus \Omega_2$ be
the relative complement of $\Omega_2$ in $\Omega_1$, i.e., 
$\Omega_1 \setminus \Omega_2 := 
\{\omega \in \Omega_1:~ \omega \not\in \Omega_2 \}$.

Let $M^{\top}$ denote the transpose of a matrix $M \in \mathbb{R}^{\sf n\times m}$.
The Euclidean norm of a vector $v \in \mathbb{R}^{\sf n}$ is defined by
$\|v\| := (v^{\top}v)^{1/2}$.
For a matrix $M \in \mathbb{R}^{\sf m\times n}$, its Euclidean induced norm is
defined by $\|M\| := \sup \{ \|Mv\|:~v\in \mathbb{R}^{\sf n},~\|v\|= 1 \}$.
Let $\lambda_{\max}(P)$ and $\lambda_{\min}(P)$ denote
the largest and the smallest eigenvalue of 
a square matrix $P \in \mathbb{R}^{\sf n\times n}$.
Let $\mathcal{B}(L)$ be
the closed ball in $\mathbb{R}^{\sf n}$ with center at the origin and radius $L$, that is,
$\mathcal{B}(L) :=
\{x \in \mathbb{R}^{\sf n}:~ \|x \| \leq L \}$.

Let $T_s$ be the sampling period. For $t \geq 0$,
we define $[t]^-$ by
\begin{equation*}
	[t]^- := kT_s\qquad \text{if~~~}kT_s \leq t < (k+1)T_s \qquad (k \in \mathbb{Z}_+).
\end{equation*}

\section{Sampled-data Switched Systems with Quantization}
\subsection{Switched systems}
Consider the following 
continuous-time switched linear system
\begin{equation}
	\label{eq:SLS}
	\dot x = A_{\sigma}x + B_{\sigma}u, 
	%~~ x(0) = x_0,~\sigma(0) = p_0,
\end{equation}
where $x(t) \in \mathbb{R}^{\sf n}$ is the state and 
$u(t) \in \mathbb{R}^{\sf m}$ is the control input.
For a finite index set $\mathcal{P}$, the mapping 
$\sigma:~[0,\infty) \to \mathcal{P}$ is right-continuous and piecewise constant,
which indicates the active mode $\sigma(t) \in \mathcal{P}$ at each time $t\geq 0$.
We call $\sigma$ a \textit{switching signal}, and the discontinuities of $\sigma$ 
\textit{switching times} or simply
\textit{switches}.
The plant sends to the controller the state $x$ 
and the switching signal $\sigma$.

The first assumption is stabilizability 
of all modes.
\begin{assum}
	\label{ass:system}
	For every mode $p \in \mathcal{P}$, $(A_p, B_p)$ is stabilizable, i.e., 
	there exists a feedback gain $K_p \in \mathbb{R}^{\sf m \times n}$ such that
	$A_p+B_pK_p$ is Hurwitz. 
\end{assum}
%Furthermore, every sampling interval has 
%at most one switch.
%\end{assum}
%Remark \ref{rem:AssumptionRem} (3) below shows that
%the reason why we need the switching assumption.

\subsection{Quantized sampled-data system}
\begin{figure}[b]
	\centering
	\includegraphics[width = 8.5cm,clip]
	{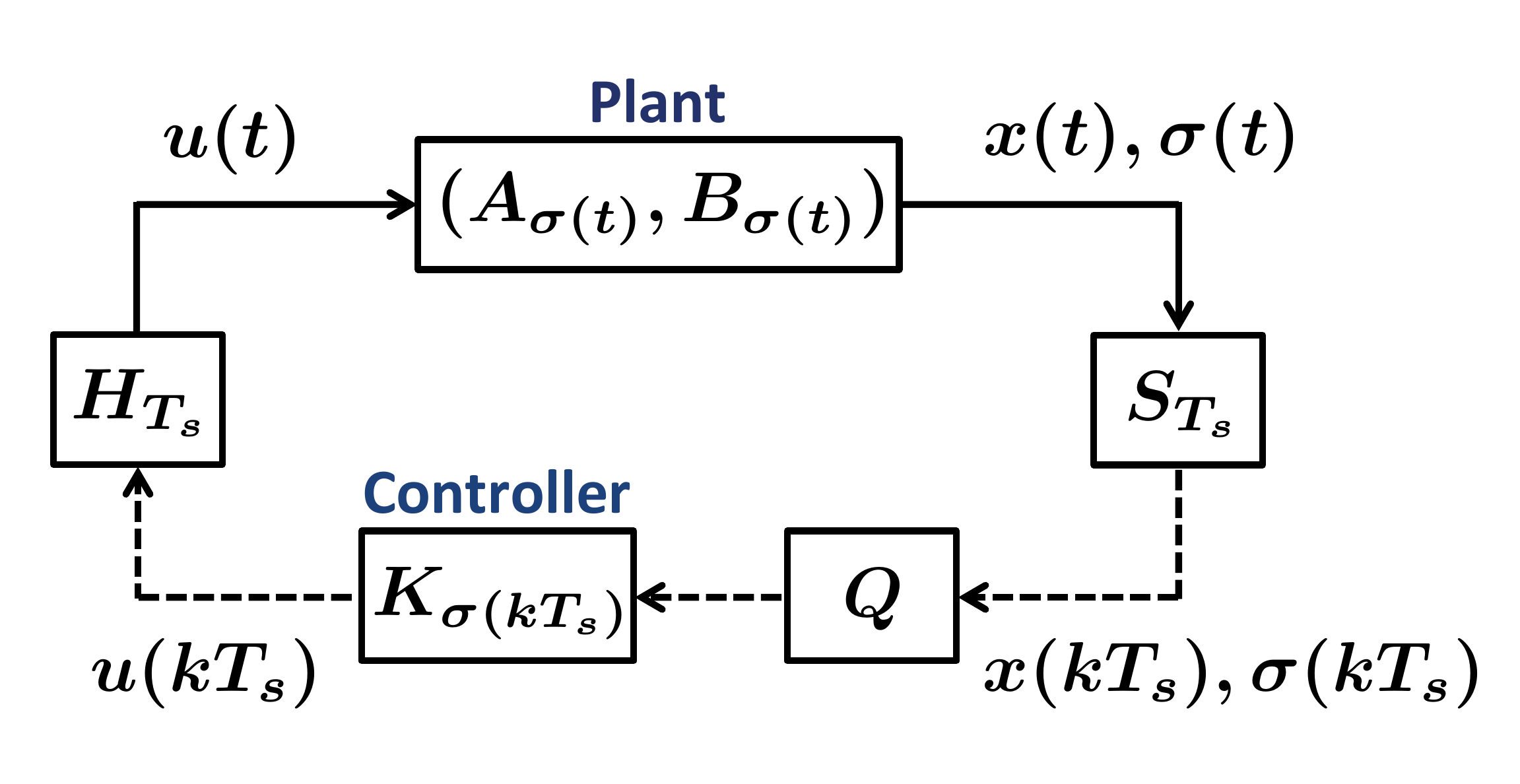}
	\caption{Sampled-data switched system with quantization, 
		where $T_s$ is the sampling period and
		$S_{T_s}$, $H_{T_s}$, and $Q$ are a sampler, a zero-order hold, and
		a static quantizer, respectively.}
	\label{fig:WSDSLS}
\end{figure}
Consider the closed-loop system in Fig.~\ref{fig:WSDSLS}.
Let $T_s > 0$ be the sampling period.
The sampler $S_{T_s}$ is given by
\begin{equation*}
	S_{T_s}: (x,\sigma) \mapsto (x(kT_s),\sigma(kT_s))\qquad (k \in \mathbb{Z}_+),
\end{equation*}
and
the zero-order hold $H_{T_s}$ by
\begin{equation*}
	H_{T_s}\!:u_d \!\mapsto\! u(t)\!=\!u_d(k),~ t\in[kT_s,(k+1)T_s)~~
	(k \in \mathbb{Z}_+).
\end{equation*}

The second assumption is that
at most one switch happens in each sampling interval.
\begin{assum}
	\label{ass:at_most_one_switch}
	Every sampling interval $(kT_s, (k+1)T_s)$ has at most one switch.
\end{assum}
See Remark \ref{rem:AssumptionRem} (3) below for
the reason why we need this switching assumption.

We now state the definition of a memoryless quantizer $Q$ given in \cite{Ishii2004}.
For an index set $\mathcal{S}$,
the partition $\{\mathcal{Q}_j\}_{j \in \mathcal{S}}$ 
of $\mathbb{R}^{\sf n}$ is 
said to be \textit{finite} if for every bounded set $B$, there exists a
finite subset $\mathcal{S}_f$ of $\mathcal{S}$ such that
$B \subset \bigcup_{j \in \mathcal{S}_f} \mathcal{Q}_j$.
We define
the quantizer $Q$ 
with respect to the finite partition $\{\mathcal{Q}_j\}_{j \in \mathcal{S}}$
by
\begin{align*}
	%\label{eq:quantizer_def}
	Q:~\mathbb{R}^{\sf n} &\to 
	\{q_j \}_{j \in \mathcal{S}} \subset \mathbb{R}^{\sf n} \\
	x &\mapsto q_j \quad \text{if~~} x \in \mathcal{Q}_j\quad (j \in \mathcal{S}).
\end{align*}

As in \cite{Liberzon2003Automatica, Liberzon2007}, 
we assume that $Q(x) = 0$ if $x$ is close to the origin: 
\begin{assum}
	\label{ass:quantization_near_origin}
	If $\Cl (\mathcal{Q}_j)$ contains the origin, then 
	the corresponding quantization value $q_j = 0$.
\end{assum}

Let $q_x$ be the output of the zero-order hold whose input is 
the quantized state at sampling times, i.e.,
$
%\label{eq:q_x_def}
q_x(t) = Q(x([t]^-)).
$
Note that in Fig. \ref{fig:WSDSLS}, the control input $u$ is given by
\begin{equation}
	\label{eq:control_input}
	%K_{\sigma([t]^-)}q_x(t) = 
	%H_{T_s} 
	%\big(
	%K_{\sigma([t]^-)} Q(x([t]^-))
	%\big).
	u(t) = K_{\sigma([t]^-)}q_x(t).
\end{equation}
The control input $u$
is a piecewise-constant and discrete-valued
signal.
If we assume that a finite subset $\mathcal{S}_f$
of $\mathcal{S}$ satisfies $x(t)
\in \bigcup_{j \in \mathcal{S}_f} \mathcal{Q}_j$
for every state trajectory $x(t)$, then
data is transmitted to/from the controller 
at the rate of
\begin{equation*}
	\frac{\log_2 |\mathcal{S}_f| + 
		\log_2 |\mathcal{P}|	}{T_s}
\end{equation*} 
bits per time unit, 
where  $|\mathcal{S}_f|$ 
and $|\mathcal{P}|$
are the numbers of elements in $\mathcal{S}_f$ and
$\mathcal{P}$, respectively.

Let $P \in \mathbb{R}^{\sf n \times n}$ be positive definite and 
define the quadratic Lyapunov function
$V(x) := x^{\top} P x$ for $x \in \mathbb{R}^{\sf n}$.
Its time derivative $\dot V$ along the trajectory of \eqref{eq:SLS} with \eqref{eq:control_input}
is given by
\begin{align}
	\dot V((t),q_x(t),\sigma(t)) 
	= (A_{\sigma(t)}x(t)+B_{\sigma(t)}K_{\sigma([t]^-)}&q_x(t))^{\top}Px(t) 
	\notag \\
	&+x(t)^{\top}P(A_{\sigma(t)}x(t)+B_{\sigma(t)}K_{\sigma([t]^-)}q_x(t))
	\label{eq:dotV_def}
\end{align}
if $t$ is not a switching time or a sampling time.

For $p,q \in \mathcal{P}$ with $p \not= q$, 
we also define $\dot V_p$ and $\dot V_{p,q}$ by
\begin{align}
	\dot V_p(x(t),q_x(t))
	&:=
	(A_px(t)+B_pK_pq_x(t))^{\top}Px(t) 
	%\label{eq:dotVp_def} 
	+x(t)^{\top}P(A_px(t)+B_pK_pq_x(t)) \notag \\
	\dot V_{p,q}(x(t),q_x(t)) 
	&:=
	(A_px(t)+B_pK_qq_x(t))^{\top}Px(t)
	+x(t)^{\top}P(A_px(t)+B_pK_qq_x(t)). 
	\label{eq:Vpq_def}
\end{align}
Then
$\dot V_{p}$ and $\dot V_{p,q}$ are the time derivatives of $V$
along the trajectories of the systems $(A_p, B_pK_p)$ and $(A_p, B_pK_q)$, 
respectively.

Every individual mode is assumed to be 
stable in the following sense with
the common Lyapunov function $V$:
\begin{assum}
	\label{ass:subsystem_QA}
	Consider 
	the following quantized sampled-data systems with `a single mode':
	\begin{equation}
		\label{eq:subsystem_QSDS}
		\dot x = A_px + B_pu, \quad u = K_pq_x \qquad (p \in \mathcal{P}).
	\end{equation}
	Let $C$ be a positive number and
	suppose that $R$ and $r$ satisfy $R > r > 0$. Then
	there exists a positive-definite matrix $P \in \mathbb{R}^{\sf n \times n}$
	such that for all $p \in \mathcal{P}$, every trajectory $x$ of the system
	\eqref{eq:subsystem_QSDS} with $x(0) \in \overline{\mathcal{E}}_P(R)$
	satisfies
	\begin{equation}
		\label{eq:dotVp_bound}
		\dot V_p(x(t),q_x(t)) \leq -C \|x(t)\|^2
	\end{equation}
	or $x(t) \in \underline{\mathcal{E}}_P(r)$ for all $t \geq 0$,
	where $\overline{\mathcal{E}}_P(R)$ and $\underline{\mathcal{E}}_P(r)$
	are given by
	\begin{align*}
		\overline{\mathcal{E}}_P(R) &:= 
		\{
		x \in \mathbb{R}^{\sf n}:~V(x) \leq R^2\lambda_{\max}(P)
		\}\\
		\underline{\mathcal{E}}_P(r) &:= 
		\{
		x \in \mathbb{R}^{\sf n}:~V(x) \leq r^2\lambda_{\min}(P)
		\}.
	\end{align*}
\end{assum}

Assumption \ref{ass:subsystem_QA} implies the following:
If we have no switches, then the
common Lyapunov function $V$ exponentially decreases
at a certain rate until
$V \leq r^2\lambda_{\min}(P)$
for every mode $p \in \mathcal{P}$. 
Furthermore, the trajectory does not leave
$\underline{\mathcal{E}}_P(r)$
as well as
$\overline{\mathcal{E}}_P(R)$ once it falls into them.

The objective of the present paper is to
find a switching condition under which
every trajectory of  the switched system in Fig.~\ref{fig:WSDSLS} 
falls into some neighborhood of the origin and remains in the neighborhood. 
We also determine how small the neighborhood is.

\begin{rem}
	\label{rem:AssumptionRem}
	%\noindent
	{\bf (1)}
		The ellipsoid $\overline{\mathcal{E}}_P(R)$ is the \textit{smallest} level set of $V$
		\textit{containing} $\mathcal{B}(R)$, whereas $\underline{\mathcal{E}}_P(r)$ is
		the \textit{largest} level set of $V$ \textit{contained in} $\mathcal{B}(r)$.

	\noindent
	%{\bf (b)}
	{\bf (2)}
		For switched systems without samplers,
		the existence of common Lyapunov functions is a sufficient condition 
		for stability under arbitrary switching; see, e.g., 
		\cite{Liberzon2003Book, Lin2009, Shorten2007}.
		For sampled-data switched systems, however,
		such functions do not guarantee stability because
		switching within a sampling interval may make the closed-loop system unstable. 
		
		\noindent
		{\bf (3)}
		Not only sampling but also quantization makes the stability analysis complicated. In fact, 
		Assumption \ref{ass:subsystem_QA} does not consider
		trajectories after a switch even without a mode mismatch.
		For example, suppose that the mode changes $p \to q \to p$ 
		at the switching times $t_1$ and $t_2$ 
		in a sampling interval $(0, T_s)$.
		Although the modes coincide between the plant and the controller
		in $[t_2, T_s)$, 
		\eqref{eq:dotVp_bound} holds only for $t \in (0,t_1)$.
		This is because the trajectory in $[t_2, T_s)$ does not
		appear for systems with a single mode. 
		In Assumption \ref{ass:at_most_one_switch}, we therefore assume that 
		at most one switch occurs in
		a sampling interval, and hence \eqref{eq:dotVp_bound} holds
		whenever the modes coincide.
		If we consider trajectories in the worst case, then 
		the switching condition in Assumption \ref{ass:at_most_one_switch} 
		can be removed. 
		However, the stability analysis becomes more conservative and involved.
		
		\noindent
		%{\bf (a)}
		{\bf (4)}
		For quantized sampled-data plants with a single mode,
		the authors of \cite{Ishii2004} have
		proposed a randomized algorithm for the computation of $P$ in 
		Assumption~\ref{ass:subsystem_QA}.
		On the other hand, for
		switched systems without sampler or quantizer, 
		the authors of \cite{Liberzon2004} have developed
		a randomized algorithm to construct
		common Lyapunov functions.
		Combining these algorithms together,
		we can efficiently compute the desired common Lyapunov function.
		See Appendix B for details of 
		the randomized algorithm.

		%
		%\noindent
		%{\bf (d)} Since we construct $P$
		%by the randomized algorithm in \cite{Ishii2004},
		%{\em we further assume that $P$ satisfies the sufficient condition for
		%quadratic stability in \cite[Proposition 3.1]{Ishii2004}.}
		%The condition is used only in Lemma \ref{lem:r_to_ar} below.
	%\end{enumerate}
\end{rem}

\section{Stabilization with Limited Information}
\subsection{Upper bound on $\dot V_{p,q}$}
Assumption~\ref{ass:subsystem_QA} gives an upper bound \eqref{eq:dotVp_bound}
on $\dot V_{p}$, i.e., 
the decreasing rate of the Lyapunov function in the case when
we use the feedback gain consistent with
the currently active mode of the plant. 
In this subsection,
we will
find an upper bound on $\dot V_{p,q}$, i.e.,
the growth rate in the case when intersample switching leads to the
mismatch of the modes between the plant and the feedback gain.
More specifically, the aim here is to obtain $D > 0$ satisfying
\begin{align}
	\label{eq:dotVpq_bound}
	\dot V_{p,q}(x(t),q_x(t)) \leq D \|x(t)\|^2.
\end{align}

Let $q_x(t) - x(t)$ is the error between
the sampled and quantized state $q_x(t)$ and 
the state $x(t)$ at the present time.
Since 
\begin{align}
	\dot V_{p,q}(x(t),q_x(t)) 
	&=
	2x(t)^{\top}P(A_p + B_pK_q)x(t) +
	2x(t)^{\top}PB_pK_q(q_x(t) - x(t)),
	\label{eq:dot_V_qx_x}
\end{align}
we need to obtain
a bound on the error $q_x(t) - x(t)$ 
by using $x(t)$.
%we shall investigate intersample switching that leads to the
%mismatch of the modes between the plant and the feedback gain, and shall
%find two upper bounds of $\dot V$ in such a case, i.e., $\dot V_{p,q}$ in \eqref{eq:Vpq_def}.
We begin by examining the relationship among 
the state at the present time $x(t)$, the sampled state $x([t]^-)$,
and the sampled quantized state $q_x(t)$.

The partition $\{\mathcal{Q}_j\}_{j \in \mathcal{S}}$ is finite. Furthermore,
Assumption \ref{ass:quantization_near_origin} shows that
if there exists a sequence $\{\xi_k\} \subset \mathcal{Q}_j$ such that
$\xi_{k} \to 0$ ($k \to \infty$),
then $Q(x) = 0$ for all $x \in \mathcal{Q}_j$.
Hence
there exists a constant $\alpha_0 > 0$ such that
\begin{align}
	\|B_pK_q Q(x)\| \leq \alpha_0 \|x\|  \label{eq:alpha0_bound}
\end{align}
for all $p,q \in \mathcal{P}$ and $x \in
\overline{\mathcal{E}}_P(R)$; see Remark \ref{rem:growth_rate} (3) for
the computation of $\alpha_0$.
We also define $\Lambda$ by 
\[
\Lambda := \max_{p \in \mathcal{P}} \|A_p\|.
\]

The next result gives an upper bound of
the norm of the sampled state $x([t]^-)$ by using
the state at the present time $x(t)$.

\begin{lem}
	\label{lem:alpha1_bound}
	Consider the swithced system \eqref{eq:SLS} with
	\eqref{eq:control_input}, where $\sigma$ has finitely many switching times
	in every finite interval.
	Suppose that
	\begin{equation}
		\label{eq:alpha_0_condition}
		\eta := \alpha_0 \frac{e^{\Lambda T_s} - 1}{\Lambda} < 1,
	\end{equation}
	and define $\alpha_1$ by
	\begin{align}
		\label{eq:alpha1_def}
		\alpha_1 := 
		\frac{e^{\Lambda T_s}}{1 - \eta}.
	\end{align}
	Then 
	we have
	\begin{align}
		\label{eq:alpha1_bound}
		\| x([t]^-)\| <
		\alpha_1 \|x(t)\|
	\end{align}
	for all $t \geq 0$ with 
	$x([t]^-) \in
	\overline{\mathcal{E}}_P(R)$.
\end{lem}
\begin{pf}
	It suffices to prove \eqref{eq:alpha1_bound} for
	$x(0) \in \overline{\mathcal{E}}_P(R)$ and 
	$t \in [0,T_s)$.
	
	Let $\Phi(\tau_1,\tau_2)$ denote the state-transition matrix of 
	the switched system \eqref{eq:SLS}
	for $\tau_1 \geq \tau_2$.
	If no switches occur, $\Phi(\tau_1,\tau_2)$ is given by 
	$\Phi(\tau_1,\tau_2) = e^{(\tau_1-\tau_2)A_{\sigma(\tau_2)}}$.
	If $t_1,t_2,\dots, t_m$ are the switching times in an interval $(\tau_2,\tau_1)$
	and if we define $t_0:=\tau_2$ and $t_{m+1}:=\tau_1$, then
	we have
	\begin{align*}
		\Phi(\tau_1,\tau_2) = 
		%e^{(\tau_1 - t_m)A_{\sigma(t_m)}} \\
		%\times
		%Assumption \ref{ass:subsystem_QA}
		\prod_{k=0}^{m} e^{(t_{k+1} - t_k)A_{\sigma(t_k)} }.
		%\cdot
		%\right)
		%e^{(t_1 - \tau_2)A_{\sigma(\tau_2)}}.
		%\label{eq:state_trans_map}
	\end{align*}
	%\begin{align*}
	%%\Phi(t,\tau) = e^{(t - t_m)A_{\sigma(t_m)}} &\cdot
	%%e^{(t_{m} - t_{m-1})A_{\sigma(t_{m-1})}} \notag \\
	%%&e^{(t_{2} - t_{1})A_{\sigma(t_{1})}} \cdot
	%%e^{(t_1 - \tau)A_{\sigma(\tau)}}
	%%\label{eq:state_trans_map}
	%\Phi(\tau_1,\tau_2) = e^{(\tau_1 - t_m)A_{\sigma(t_m)}}
	%\cdot
	%%\times
	%%Assumption \ref{ass:subsystem_QA}
	%\prod_{k=1}^{m-1} e^{(t_{k+1} - t_k)A_{\sigma(t_k)} }
	%%\right)
	%\cdot
	%e^{(t_1 - \tau_2)A_{\sigma(\tau_2)}}.
	%%\label{eq:state_trans_map}
	%\end{align*}

	Since
	\begin{equation}
		\label{eq:state_t}
		x(t) = \Phi(t,0) x(0) + \int^{t}_0 \Phi(t,\tau)
		B_{\sigma(\tau)} K_{\sigma(0)} q_x(\tau) d \tau
	\end{equation}
	and since $\Phi(\tau,0)^{-1} = \Phi(t,0)^{-1}\Phi(t,\tau)$, it follows that
	\begin{equation*}
		x(0) = \Phi(t,0)^{-1} x(t) + \int^{t}_0 \Phi(\tau,0)^{-1}
		B_{\sigma(\tau)} K_{\sigma(0)} q_x(\tau) d \tau.
	\end{equation*}
	This leads to
	\begin{align}
		\|x(0)\| \leq &\|\Phi(t,0)^{-1}\| \cdot \|x(t)\| + \left\| \int^{t}_0 \Phi(\tau,0)^{-1}
		B_{\sigma(\tau)} K_{\sigma(0)} q_x(\tau) d \tau \right\|.
		\label{eq:x0_bound}
	\end{align}
	Let $t_1,t_2,\dots, t_m$ be the switching times in the interval $[0,t)$.
	Since $\| e^{\tau A}\| \leq e^{\tau\|A\|}$ for $\tau \geq 0$,
	if we define $t_0:=0$ and $t_{m+1} := t$, then
	we obtain
	\begin{align}
		\|\Phi(t,0)^{-1} \|  
		%&~~\leq
		%\| e^{-t_1A_{\sigma(0)}} \| \cdot 
		%\prod_{k=1}^{m-1}
		%\| e^{-(t_{k+1} - t_{k})A_{\sigma(t_{k})}}\| \cdot
		%\| e^{-(t - t_m)A_{\sigma(t_m)}}\| \notag \\
		\leq
		%e^{(t - t_m) \| A_{\sigma(t_m)} \| }  \cdot
		%\prod_{k=1}^{m-1}
		%e^{(t_{k+1} - t_{k}) \| A_{\sigma(t_{k})} \| } \cdot
		%e^{t_1\| A_{\sigma(0)} \|} \notag \\
		\prod_{k=0}^{m}
		e^{(t_{k+1} - t_{k}) \| A_{\sigma(t_{k})} \| }
		\leq
		e^{\Lambda t} < e^{\Lambda T_s} \label{eq:state_map_bound}.
	\end{align}
	It is obvious that the equation above holds in the non-switched case as well.
	Since $q_x(\tau) = q_x(0) = Q(x(0))$ for
	all $\tau \in [0,T_s]$, 
	if follows from \eqref{eq:alpha0_bound} that
	\begin{align}
		\left\| \int^{t}_0 \Phi(\tau,0)^{-1} 
		B_{\sigma(\tau)} K_{\sigma(0)} q_x(\tau) d \tau \right\|
		& \leq
		\int^{t}_0
		\|\Phi(\tau, 0)^{-1} \| \cdot \|B_{\sigma(\tau)} K_{\sigma(0)} q_x(\tau) \| d\tau 
		\notag \\
		&\leq
		\alpha_0 \int^{t}_0 e^{\Lambda \tau} d\tau  \|x(0)\| \notag \\
		&\leq
		\alpha_0  \frac{e^{\Lambda T_s} - 1}{\Lambda}  \|x(0)\|
		= \eta \|x(0)\|.
		\label{eq:int_state_map_bound}
	\end{align}
	Substituting \eqref{eq:state_map_bound} and \eqref{eq:int_state_map_bound}
	into \eqref{eq:x0_bound},
	we obtain
	\begin{equation*}
		\|x(0)\| < e^{\Lambda T_s} \|x(t)\| + 
		\eta \|x(0)\|.
	\end{equation*}
	Thus if \eqref{eq:alpha_0_condition} holds, 
	\eqref{eq:alpha1_bound} follows.
\end{pf}

Let us next develop an upper bound of the norm of
the error $x(t) - x([t]^-)$ due to sampling.
To this end, 
we use the following property of the state-transition map of a switched system:
\begin{prop}
	\label{prop:state_transition_bound}
	Let $\Phi(t,0)$ be the state-transition map of 
	the switched system \eqref{eq:SLS}
	as above. Then
	\begin{equation}
		\label{eq:Phi_1_diff_bound}
		\|\Phi(t,0) - I\| \leq e^{\Lambda t} - 1.
	\end{equation}
\end{prop}
%\begin{pf}[Outline only]
%If no switch occurs, we immediately obtain \eqref{eq:Phi_1_diff_bound}
%by using the partial sum of $\Phi(t,0) = e^{tA_{\sigma(0)}}$.
%The general case follows by induction and the following inequality:
%\[
%\|e^{t_1A_1}e^{t_2A_2} - I\|
%\leq
%\|e^{t_1A_1} - I\| \cdot \|e^{t_2A_2}\|
%+ \|e^{t_2A_2} - I\|.
%\]
%\end{pf}

\begin{pf}
	Let us first consider the case without switching; that is,
	\begin{equation}
		\label{eq:no_switching_matrix_expo_diff_bound}
		\|e^{tA_{\sigma(0)}} - I \| \leq e^{\Lambda t} - 1.
	\end{equation}
	Define the partial sum $S_N$ of $e^{tA_{\sigma(0)}} - I$ by
	\begin{equation*}
		S_N(t) := 
		\sum_{k=0}^N \frac{1}{k!} (tA_{\sigma(0)})^k - I
		=
		\sum_{k=1}^N \frac{1}{k!} (tA_{\sigma(0)})^k.
	\end{equation*}
	Then for all $t \geq 0$, we have
	\begin{align*}
		\|S_N(t)\| &\leq
		\sum_{k=1}^N \frac{1}{k!} \left(t\|A_{\sigma(0)}\| \right)^k  \\
		&=\sum_{k=0}^N \frac{1}{k!} \left(t\|A_{\sigma(0)}\|\right)^k  - 1\\
		&\leq \sum_{k=0}^\infty \frac{1}{k!} \left(t\|A_{\sigma(0)}\|\right)^k  - 1\\
		&= e^{t\|A_{\sigma(0)}\|} - 1 
		\leq e^{\Lambda t} - 1.
	\end{align*}
	Letting $N \to \infty$, 
	we obtain \eqref{eq:no_switching_matrix_expo_diff_bound}.
	
	We now prove \eqref{eq:Phi_1_diff_bound}
	in the switched case.
	Let $t_1,t_2,\dots,t_m$ be the switching times in the interval $(0,t)$.
	Let $t_0 = 0$ and $t_{m+1} = t$.
	Then \eqref{eq:Phi_1_diff_bound} is equivalent to
	\begin{align}
		\left\|~
		\prod_{k=0}^{m} 
		e^{(t_{k+1} - t_{k})A_{\sigma(t_{k})}}- I~
		\right\| 
		\leq e^{\Lambda t} - 1.
		\label{eq:matrix_expo_diff_bound}
	\end{align}
	
	We have already shown 
	\eqref{eq:matrix_expo_diff_bound} in the case $m=0$, i.e., the non-switched case.
	The general case follows by induction. For $m \geq 1$,
	\begin{align*}
		&\left\| \prod_{k=0}^{m} 
		e^{(t_{k+1} - t_{k})A_{\sigma(t_{k})}}- I 
		\right\| \\
		&\qquad \leq
		\left\| e^{(t_{m+1} - t_{m})A_{\sigma(t_{m})}} \left(
		\prod_{k=0}^{m-1} e^{(t_{k+1} - t_{k})A_{\sigma(t_{k})}}
		- I \right) \right\|  + 
		\| e^{(t_{m+1} - t_{m})A_{\sigma(t_{m})}}- I \| \\
		&\qquad \leq\| e^{(t_{m+1} - t_{m})A_{\sigma(t_{m})}}\| \cdot
		\left\|
		\prod_{k=0}^{m-1}
		e^{(t_{k+1} - t_{k}) A_{\sigma(t_{k})} } - I
		\right\|  + 
		\| e^{(t_{m+1} - t_{m})A_{\sigma(t_m)}}- I \|.
	\end{align*}
	Hence if \eqref{eq:matrix_expo_diff_bound} holds with
	$m-1$ in place of $m$, then
	\begin{align*}
		&\| e^{(t_{m+1} - t_{m})A_{\sigma(t_{m})}}\| \cdot
		\left\|
		\prod_{k=0}^{m-1}
		e^{(t_{k+1} - t_{k}) A_{\sigma(t_{k})} } \!-\! I
		\right\|  + 
		\| e^{(t_{m+1} - t_{m})A_{\sigma(t_m)}} \!-\! I \| \\
		&\qquad \leq
		e^{\Lambda (t_{m+1} - t_{m})}
		(e^{\Lambda t_{m}} \!-\! 1) + (e^{\Lambda (t_{m+1} - t_{m})} \!-\! 1) \\
		&\qquad = e^{\Lambda t } - 1.
	\end{align*}
	Thus we obtain \eqref{eq:matrix_expo_diff_bound}.
\end{pf}

\begin{lem}
	\label{lem:beta1_bound}
	Consider the switched system \eqref{eq:SLS} with \eqref{eq:control_input}, 
	where $\sigma$ has finitely many switching times
	in every finite interval.
	Define $\beta_1$ by
	\begin{align}
		\label{eq:beta1_def}
		\beta_1 := (e^{\Lambda T_s} - 1) \left( 1 +  \frac{\alpha_0}{\Lambda} \right)
	\end{align}
	Then we have
	\begin{equation}
		\label{eq:beta1_bound}
		\|x(t) - x([t]^-) \| < \beta_1 \|x([t]^-)\|
	\end{equation}
	for all $t \geq 0$ with 
	$x([t]^-) \in
	\overline{\mathcal{E}}_P(R)$.
\end{lem}
%Lemma \ref{lem:beta1_bound} follows immediately from 
%Proposition \ref{prop:state_transition_bound},
%so its proof is omitted for reason of space.
\begin{pf}
	As in the proof of Lemma \ref{lem:alpha1_bound}, it suffices to prove
	\eqref{eq:beta1_bound} for all 
	$x(0) \in \overline{\mathcal{E}}_P(R)$ and 
	$t \in [0,T_s)$.
	
	By \eqref{eq:state_t}, we obtain
	\begin{align*}
		x(t) - x(0) = 
		(\Phi(t,0) &- I) x(0)  
		+ \int^{t}_0 \Phi(t,\tau)
		B_{\sigma(\tau)} K_{\sigma(0)} q_x(\tau) d \tau.
	\end{align*}
	This leads to
	\begin{align}
		\|x(t) - x(0)\| \leq
		&\| \Phi(t,0) - I \| \cdot \|x(0) \| + \left\| \int^{t}_0 \Phi(t,\tau)
		B_{\sigma(\tau)} K_{\sigma(0)} q_x(\tau) d \tau \right\|.
		\label{eq:xt_x0_dif_bound}
	\end{align}
	
	Proposition \ref{prop:state_transition_bound} provides
	the following upper bound on the first term of the right-hand side of 
	\eqref{eq:xt_x0_dif_bound}:
	\begin{align}
		\label{eq:first_term_upper}
		\|\Phi(t,0) - I\| \leq
		e^{\Lambda t} - 1 < e^{\Lambda T_s} - 1.
	\end{align}
	Since 
	%a
	%calculation similar to \eqref{eq:state_map_bound} shows that
	$\| \Phi (t,\tau) \| \leq e^{\Lambda(t-\tau)}$,
	a calculation similar to \eqref{eq:int_state_map_bound}
	gives
	\begin{align}
		\left\| 
		\int^{t}_0 \Phi(t,\tau)
		B_{\sigma(\tau)} K_{\sigma(0)} q_x(\tau) d \tau \right\|
		%&\leq
		%\alpha_0 \int^{t}_0 e^{\Lambda (t - \tau)} d \tau \cdot
		%\|x(0)\| \notag \\
		&\leq
		\alpha_0 \frac{e^{\Lambda T_s} - 1}{\Lambda} \|x(0)\|.
		\label{eq:int_Phit_tau}
	\end{align}
	We obtain \eqref{eq:beta1_bound} by
	substituting \eqref{eq:first_term_upper}
	and \eqref{eq:int_Phit_tau} into \eqref{eq:xt_x0_dif_bound}.
\end{pf}

We are now in the position to obtain an upper bound of
the norm of  
%the
%sampled quantized state $q_x(t)$ and that of 
the error 
$q_x(t) - x(t)$ due to
sampling and quantization by using
the original state $x(t)$.

Similarly to \eqref{eq:alpha0_bound},
to each $p,q \in \mathcal{P}$ with $p\not=q$, 
there corresponds a positive number
$\gamma_0(p,q)$ such that
\begin{align}
	%\label{eq:alpha2_bound}
	%\|PB_p(K_q- K_p)Q(x)\| &\leq \alpha_2(p,q)  \|x\| \\
	\label{eq:beta2_bound}
	\|PB_pK_q(Q(x) - x)\| &\leq \gamma_0(p,q) \|x\|
\end{align}
for all $x \in
\overline{\mathcal{E}}_P(R)$; see Remark \ref{rem:growth_rate} (3) for
the computation of $\gamma_0$.

%the inequalities \eqref{eq:alpha_bound} and 
%\eqref{eq:beta_bound} below.
\begin{lem}
	\label{thm:alpha_beta_bound}
	Consider the switched system \eqref{eq:SLS} with \eqref{eq:control_input}, 
	where $\sigma$ has finitely many switching times
	in every finite interval.
	Define $\alpha_1$ and $\beta_1$ as in Lemmas 
	\ref{lem:alpha1_bound} and \ref{lem:beta1_bound}.
	If $\gamma(p,q)$ is defined by
	\begin{align}
		\label{def:gam_pq}
		%\alpha(p,q) &= \alpha_1\alpha_2(p,q)\\
		\gamma(p,q) &:= \alpha_1 (\beta_1\|PB_pK_q\|  + \gamma_0(p,q))
	\end{align}
	for each $p,q \in \mathcal{P}$ with $p\not=q$,
	then $\gamma(p,q)$ satisfies
	\begin{align}
		%\|PB_p(K_q- K_p)q_x(t)\| &< \alpha (p,q) \|x(t)\| \label{eq:alpha_bound} \\
		\|PB_pK_q(q_x(t) - x(t))\| &< \gamma(p,q)\|x(t)\| \label{eq:beta_bound}
	\end{align}
	for all $t \geq 0$ with 
	$x([t]^-) \in
	\overline{\mathcal{E}}_P(R)$.
\end{lem}
\begin{pf}
	%We obtain
	%the first inequality \eqref{eq:alpha_bound} by
	%\eqref{eq:alpha1_bound} and \eqref{eq:alpha2_bound}.
	Since $q_x(t) = Q(x([t]^{-}))$,
	it follows from \eqref{eq:beta1_bound} and \eqref{eq:beta2_bound} that
	\begin{align*}
		\|PB_pK_q(q_x(t) - x(t))\|  
		&\leq 
		\|PB_pK_q(q_x(t) - x([t]^-)) \| 
		 + \|PB_pK_q\|\cdot\| x([t]^-) - x(t) \| \\
		& <
		(\beta_1\|PB_pK_q\| + \gamma_0(p,q)) \|x([t]^-)\| \\
		& < \alpha_1 (\beta_1\|PB_pK_q\| + \gamma_0(p,q))\|x(t)\|.
	\end{align*}
	Thus the desired inequality \eqref{eq:beta_bound} holds.
\end{pf}

Finally, the following theorem gives the growth rate of $V$ in the case 
when the modes of the plant and the controller
are not synchronized.
\begin{thm}
	Consider the switched system \eqref{eq:SLS} with \eqref{eq:control_input}, 
	where $\sigma$ has finitely many switching times
	in every finite interval.
	Using $\gamma(p,q)$ in \eqref{def:gam_pq}, 
	we define $D$ by
	\begin{align}
		D := 
		2\max_{p \not= q} 
		(\|P(A_p + B_pK_q) \| + \gamma(p,q)).
		\label{eq:D_def}
	\end{align}
	Then
	\eqref{eq:dotVpq_bound} holds 
	for every $p,q \in \mathcal{P}$ with $p \not= q$ and
	for every $t \geq 0$ with
	$x([t]^-) \in
	\overline{\mathcal{E}}_P(R)$.
\end{thm}
%First,
%since $V_{p,q}$ satisfies
%\begin{align*}
%\dot V_{p,q}(x(t),q_x(t)) = 
%\dot V_{p}&(x(t),q_x(t))  \\
%&+ 
%2x(t)^{\top} P B_p(K_q-K_p)q_x(t),
%\end{align*}
%%\begin{align*}
%%\dot V_{p,q}(x(t),q_x(t)) = 
%%\dot V_{p}&(x(t),q_x(t))  \\
%%&+ 
%%(B_p(K_q-K_p)q_x(t))^{\top} P x(t) \\
%%&+ x(t)^{\top} P B_p(K_q-K_p)q_x(t),
%%\end{align*}
%it follows from 
%\eqref{eq:dotVp_bound} and Theorem \ref{thm:alpha_beta_bound} that
%\begin{equation}
%\label{eq:first_bound_Vpq}
%\dot V_{p,q}(x(t),q_x(t)) \leq
%(2\alpha(p,q) - C) \|x(t)\|^2
%\end{equation}
%for all $t \geq 0$ with 
%$x(t)\in
%\overline{\mathcal{E}}_P(R)~\backslash~\underline{\mathcal{E}}_P(r)$
%and $x([t]^-) \in
%\overline{\mathcal{E}}_P(R)$.
%%Here we used 
%%\begin{equation}
%%\label{eq:x_set_x_sample_set}
%%\{
%%x(t):~t\geq 0
%%\} \supset
%%\{
%%x([t]^-):~t\geq 0
%%\}.
%%\end{equation}
%The first bound \eqref{eq:first_bound_Vpq} can be negative
%if the sampling period $T_s$ and the gain difference $\|K_p - K_q\|$ are
%sufficiently small.
\begin{pf}
	Since $\dot V_{p,q}$ satisfies
	\eqref{eq:dot_V_qx_x},
	%\begin{align*}
	%\dot V_{p,q}(x(t),q_x(t)) 
	%&=
	%x(t)^{\top}(A_p + B_pK_q)^{\top}Px(t) \\
	%&\qquad+ 
	%x(t)^{\top}P(A_p + B_pK_q)x(t)  \\
	%&\qquad+
	%(q_x(t) - x(t))^{\top}K_q^{\top}B_p^{\top}Px(t) \\
	%&\qquad +
	%x(t)^{\top}PB_pK_q(q_x(t) - x(t)),
	%\end{align*}
	Lemma \ref{thm:alpha_beta_bound} shows that
	\begin{equation}
		\label{eq:second_bound_Vpq}
		\dot V_{p,q}(x(t),q_x(t)) \leq
		2(\|P(A_p + B_pK_q) \| + \gamma(p,q) ) \|x(t)\|^2
	\end{equation}
	for all $p,q \in \mathcal{P}$ with $p \not= q$ and
	for all $t \geq 0$ with 
	$x([t]^-) \in
	\overline{\mathcal{E}}_P(R)$.
	% Using \eqref{eq:x_set_x_sample_set} again, 
	Thus we obtain the desired result \eqref{eq:dotVpq_bound}.
\end{pf}

%We assume that
%{\em
%$D$ in \eqref{eq:D_def} satisfies $D \geq 0$.
%}
%This assumption involves no loss of generality. In fact,
%if $D < 0$, then $\dot V$ in \eqref{eq:dotV_def} is negative for all $\sigma$.
%This implies that
%every trajectory with its initial state in $\overline{\mathcal{E}}_P (R)$ 
%goes into $\underline{\mathcal{E}}_P (r)$ and remains there for all
%switching signals.
%Hence the stabilization of the system \eqref{eq:SLS} with \eqref{eq:control_input}
%can be achieved 
%without any information about switching signals.
%Thus the problem to be posed is trivial. 

\begin{rem}
	\label{rem:growth_rate}
	{\bf (1)}
		Fine quantization and fast sampling make
		$\alpha_1$ in \eqref{eq:alpha1_def},
		$\beta_1$ in \eqref{eq:beta1_def},
		and $\gamma_0(p,q)$ in 
		\eqref{eq:beta2_bound} small, which leads to a decrease of $D$ in \eqref{eq:D_def}.
	
	{\bf (2)}
		In this subsection,
		we have assumed that 
		finitely many switches occurs 
		in a sampling interval, which
		makes \eqref{eq:state_map_bound},
		\eqref{eq:first_term_upper}, and \eqref{eq:int_Phit_tau} conservative.
		If we allow a higher computational cost, then
		another possibility of $\alpha_1$ in \eqref{eq:alpha1_def}
		and 
		$\beta_1$ in \eqref{eq:beta1_def}
		under Assumption \ref{ass:at_most_one_switch}
		would be
		\begin{align*}
			\alpha_1 &\!=\!
			\max_{p\not=q} \max_{0 \leq t \leq T_s} \max_{0 \leq t' \leq t} 
			\frac{\| e^{-A_p t'} e^{-A_q(t - t') }\|}
			{1- \alpha_0 
				\left( \int^{t}_{t'} \|e^{-A_p t'} e^{-A_q(\tau - t')} \|  d\tau + 
				\int^{t'}_{0} \|e^{-A_p \tau}  \|  d\tau
				\right)
			}\\
			\beta_1 &\!=\!
			\max_{p\not=q} \max_{0\leq t \leq T_s} \max_{0\leq t' \leq t}
			\Biggl( \!
			\|e^{A_q (t-t')} e^{A_p t' } \!-\! I \| 
			+\!
			\alpha_0
			\Biggl(
			\int^t_{t'}
			\|e^{A_q (t-\tau)} \| d\tau 
			+\!\!
			\int^{t'}_0
			\|
			e^{A_q (t-t')} e^{A_p (t' - \tau)} d\tau
			\|
			\Biggr)\!\!
			\Biggr)\!,
		\end{align*}
		%another possibility of a bound on $\|\Phi(t,0)^{-1}\|$, $\|\Phi(t,0) - I\|$, and
		%$\|\Phi(t,\tau)\|$
		%under Assumption \ref{ass:system}
		%would be
		%\begin{align*}
		%\|\Phi(t,0)^{-1}\| 
		%&\leq \max_{p \not=q} 
		%\max_{0\leq t' \leq t}
		%\left\|
		%e^{-Aq(t-t')} \cdot e^{-A_pt'}
		%\right\|  \\
		%\|\Phi(t,0) - I\|
		%&\leq \max_{p \not=q} 
		%\max_{0\leq t' \leq t}
		%\left\|
		%e^{Aq(t-t')} \cdot e^{A_pt'}  - I
		%\right\| \\
		%\|\Phi(t,\tau)\| 
		%&\leq \max_{p \not=q} 
		%\max_{0\leq t' \leq t}
		%\left\|
		%e^{A_q(t-t')}\cdot e^{Ap (t' - \tau)}
		%\right\|.
		%\end{align*}
		where $t'$ is a switching time in $[0,t]$.
	
	\noindent
	{\bf (3)}
		We can derive $\alpha_0$ in \eqref{eq:alpha0_bound}
		and 
		$\gamma_0(p,q)$ in \eqref{eq:beta2_bound}
		as follows.
		Let $\mathcal{S}_f$
		be a subset of $\mathcal{S}$ such that
		$\overline{\mathcal{E}}_P(R) 
		\subset \bigcup_{j \in \mathcal{S}_f} \mathcal{Q}_j$.
		Then 
		\begin{equation*}
			\alpha_0 :=
			\max_{p,q \in \mathcal{P}}
			\max_{j \in \mathcal{S}_f}
			\frac{\| B_pK_q q_j\|}
			{\min_{x\in \mathcal{Q}_j} \|x\|}
		\end{equation*}
		satisfies \eqref{eq:alpha0_bound}.
		Note that
		if $\mathcal{Q}_j$ is a polyhedron, then
		$\min_{x \in \mathcal{Q}_j} \|x\|$ can be computed by
		quadratic programming; see, e.g, \cite{Boyd2004}.
		%the minimum value of the norm of a point in the quantization region 
		As regards $\gamma_0(p,q)$ in \eqref{eq:beta2_bound},
		define $\mathcal{S}_0 :=
		\{j \in \mathcal{S}:~0 \in \Cl (\mathcal{Q}_j)\}$.
		Since $Q(x) = 0$ for $x \in \mathcal{Q}_j$ with
		$j \in \mathcal{S}_0$ by Assumption \ref{ass:quantization_near_origin}, 
		it follows that $\gamma_0(p,q) \geq \|PB_pK_q\|$.
		On the other hand,
		for $j \not\in \mathcal{S}_0$, 
		we define $\hat \gamma_0$ by
		\begin{equation*}
			\hat \gamma_0(p,q) :=
			\max_{j \in \mathcal{S}_f \setminus \mathcal{S}_0}
			\frac{\|PB_pK_q\|\cdot  \max_{x \in \mathcal{Q}_j} \|q_j - x\| }
			{\min_{x \in \mathcal{Q}_j} \|x\|}.
		\end{equation*}
		Since
		\begin{align*}
			\frac{\|PB_pK_q\|\cdot  \max_{x \in \mathcal{Q}_j} \|q_j - x\| }
			{\min_{x \in \mathcal{Q}_j} \|x\|} 
			&\geq
			\frac{\|PB_pK_q\| \cdot \|q_j - x\| }{\|x\|} \\
			&\geq
			\frac{\|PB_pK_q(q_j - x)\| }{\|x\|},
		\end{align*}
		$
		\gamma_0(p,q) :=
		\max\{
		\|PB_pK_q\|,~
		\hat \gamma_0(p,q)
		\}
		$ satisfies \eqref{eq:beta2_bound}.
		We can easily compute
		$\max_{x \in \mathcal{Q}_j} \|q_j - x\| $
		if $\mathcal{Q}_j$ is a cuboid and $q_j$ is 
		a center of a vertex of $\mathcal{Q}_j$.
		In fact,
		let the set of the vertices of $\mathcal{Q}_j$ be $\mathcal{V}_j$.
		Then 
		$\max_{x \in \mathcal{Q}_j} \|q_j - x\| =
		\max_{x \in \mathcal{V}_j} \|q_j - x\|$,
		which implies that 
		$\max_{x \in \mathcal{Q}_j} \|q_j - x\|$
		can be obtained by calculating 
		$\|q_j - v\|$ for all $v \in \mathcal{V}_j$.

		%We can compute $\alpha_0$ and $\gamma_0(p,q)$ in 
		%\eqref{eq:alpha0_bound} and \eqref{eq:beta2_bound}
		%if we use a uniform quantizer or a logarithmic quantizer.
		%Let $\mathcal{S}_0$ be the set of $j$ satisfying $0 \in \Cl(\mathcal{Q}_j)$ and
		%$\mathcal{S}_1$ be the set of $j$ satisfying $0 \not\in \Cl(\mathcal{Q}_j)$, respectively.
		%Set $q_j = 0$ if $j \in \mathcal{S}_0$,
		%and let $q_j$ be the center of the cell $\mathcal{Q}_j$ if $j \in \mathcal{S}_1$.
		%For $x \in \mathcal{Q}_j$ with $j \in \mathcal{S}_0$, $Q(x) = 0$ and this leads to
		%$\gamma_0(p,q) \geq \|PB_pK_q\|$.
		%As regards $j \in \mathcal{S}_1$,
		%we only have to consider $x$ at vertices for $\alpha_0$ and
		%$x$ on faces for $\gamma_0(p,q)$, respectively.
		%If we allow some conservatism, then we can calculate $\bar{\gamma}_0(p,q)$ satisfying
		%\begin{align*}
		%\|PB_pK_q\| \cdot \|Q(x) - x\| &\leq \bar{\gamma}_0(p,q) \|x\|,
		%\end{align*}
		%instead of $\gamma(p,q)$.
		%Clearly, 
		%$\bar{\gamma}_0(p,q) \geq \gamma_0(p,q)$ and
		%$x$ only at vertices is needed for $\bar{\gamma}_0(p,q)$.
\end{rem}

\subsection{Stability analysis with total mismatch time}
Let us analyze the stability of 
the switched system \eqref{eq:SLS} with \eqref{eq:control_input} 
by using the two
upper bounds \eqref{eq:dotVp_bound} and 
\eqref{eq:dotVpq_bound} of $\dot V$.
Note that
the former bound \eqref{eq:dotVp_bound} 
is for the case $\sigma(t)=\sigma([t]^-)$, while
the latter \eqref{eq:dotVpq_bound} for the case $\sigma(t)\not=\sigma([t]^-)$.
As in \cite{Ma2015} for switched systems
with time delays,
it is therefore useful to characterize
switching signals by asynchronous periods.
\begin{defn}
	For $\tau_1 > \tau_2 \geq 0$, 
	we define the total mismatch time $\mu(\tau_1,\tau_2)$ by
	the time in which the modes mismatch between the plant and the controller, that is,
	\begin{align}
		\mu(\tau_1, \tau_2) := \text{ the length
				of the set~} \{\tau \in [\tau_2,\tau_1):\sigma(\tau) \not = \sigma([\tau]^-)\}.
		\label{eq:mu_def_same}
	\end{align}
\end{defn}
More explicitly, 
the length of a set in $\mathbb{R}$ means its Lebesgue measure.
We shall not, however, use any measure theory because
$\sigma$ has only finitely many discontinuities
in every interval.
We see that if the total mismatch time is small on average as the
average dwell-time condition introduced in \cite{Hespanha1999CDC},
then the system is stable.
We also derive an ultimate bound on the state
trajectories by using this characterization
of switching signals.

Define $C_P$ and $D_P$ by
\begin{equation*}
	C_P := \frac{C}{\lambda_{\max}(P)}, \quad
	D_P := \frac{D}{\lambda_{\min}(P)}.
\end{equation*} 

The objective of this subsection is to prove the following theorem:
\begin{thm}
	\label{thm:main_thm}
	Let Assumptions \ref{ass:system},
	\ref{ass:at_most_one_switch}, \ref{ass:quantization_near_origin}, and
	\ref{ass:subsystem_QA} hold.
	Suppose that
	$L \geq 0$ satisfies
	\begin{equation}
		\label{eq:L_inequality1}
		L < \frac{C_P}{C_P + D_P},
	\end{equation}
	and that $\kappa > 1$ satisfies
	\begin{equation}
		\label{eq:a_cond}
		\kappa^2r^2 \lambda_{\min}(P)
		< R^2\lambda_{\max}(P).
	\end{equation}
	Define $f(\kappa)$ by
	\begin{equation}
		\label{eq:ba_bound}
		f(\kappa) :=
		\frac{2\log \kappa}{C_P + D_P}.
	\end{equation}
	
	If
	$\mu$ in \eqref{eq:mu_def_same} satisfies
	\begin{equation}
		\label{eq:mu_inequality1}
		\mu(t,0) \leq Lt
	\end{equation} for every $t > 0$, and for each $T_0 \geq 0$ with 
	$\sigma(T_0) \not= \sigma([T_0]^-)$
	\begin{equation}
		\label{eq:mu_b_L_condition}
		\mu(t,T_0) \leq f(\kappa) + L(t-T_0)
	\end{equation} 
	for every $t > T_0$, then
	there exists $T_{r} \geq 0$ such that 
	for each $x(0) \in \Int(\overline{\mathcal{E}}_P(R))$ and 
	$\sigma(0) \in \mathcal{P}$,
	$x(t) \in \Int(\underline{\mathcal{E}}_P(\kappa r))$ 
	for all $t \geq T_{r}$.
	Furthermore, $x(t) \in \Int(\overline{\mathcal{E}}_P(R))$ for all $t \geq 0$.
\end{thm}

\begin{rem}
	{\bf (1)}
		Theorem \ref{thm:main_thm} gives the stability analysis of the switched system
		by using the total mismatch time of the modes 
		between the plant and the feedback gain.
		If a mismatch {\em does} occur, 
		the closed-loop system may be {\em unstable}; otherwise 
		it is {\em stable}.
		Our proposed method is therefore similar to that in \cite{Zhai2001}, where
		the stability analysis of switched systems with stable and unstable subsystems 
		is discussed with the aid of 
		the total activation time ratio between stable subsystems and unstable ones.
		In \cite{Zhai2001}, the average dwell time \cite{Hespanha1999CDC}
		is also required to be sufficiently large.
		However, such a condition is not 
		needed here because we use a common Lyapunov function.
		Conditions on the total activation time 
		ratio has been used for nonlinear systems in 
		\cite{Muller2012, Munoz2008, Yang2015}.
		Moreover, this switching characterization
		has been applied to 
		stabilization of systems with control inputs missing in \cite{Zhang2010}
		and to resilient control under denial-of-service attacks in \cite{Persis2014}.

	{\bf (2)}
		Although Theorem \ref{thm:main_thm}
		requires that \eqref{eq:mu_b_L_condition}
		holds for each $T_0 \geq 0$ with
		$\sigma(T_0) \not= \sigma([T_0]^{-})$, 
		it is enough to verify \eqref{eq:mu_b_L_condition} only
		with the sampling instant $[T_0]^- + T_s$ 
		in place of $T_0$.
		In fact, since at most one switch occurs in 
		$[[T_0]^-, [T_0]^-+T_s)$, it follows that
		if $\sigma(T_0) \not= \sigma([T_0]^-)$,
		then
		\[\mu(t, [T_0]^-+T_s) = 
		\mu(t, T_0) - ([T_0]^-+T_s - T_0).
		\]
		Hence \eqref{eq:mu_b_L_condition}
		holds for $t > T_0$ if it does for
		$t \geq [T_0]^-+T_s$. 
\end{rem}

\begin{figure}[t]
	\centering
	\includegraphics[width = 8.5cm, clip]{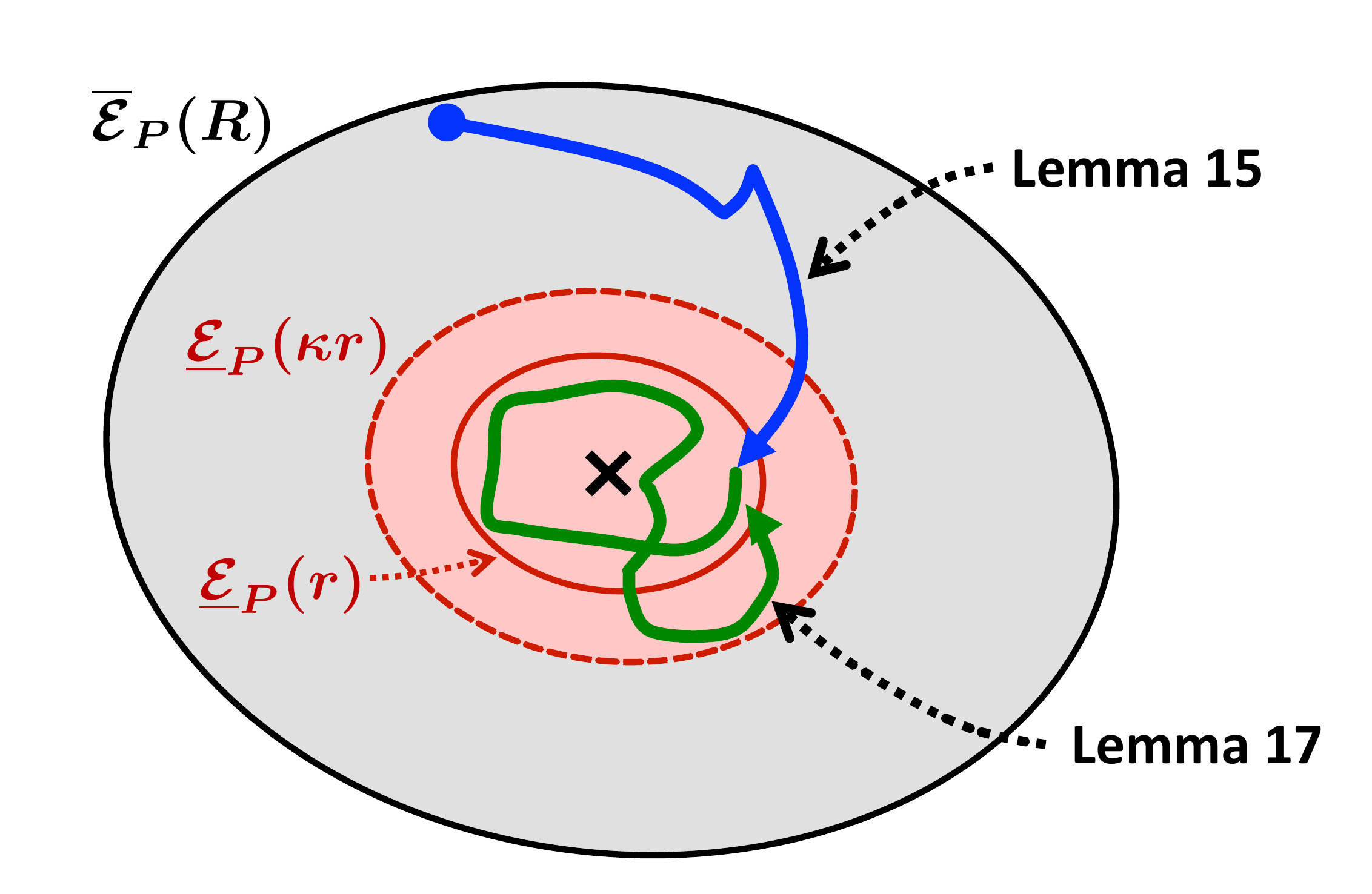}
	\caption{Behavior of trajectory}
	\label{fig:Lemma_Explain}
\end{figure}

First we study the state behavior that is outside of
$\underline{\mathcal{E}}_P(r)$.
The following lemma shows that every trajectory whose initial state is in
$\Int(\overline{\mathcal{E}}_P(R))$ falls into $\underline{\mathcal{E}}_P(r)$
%if $\mu$ satisfies an inequality determined by $C_P$ and $D_P$:
if the total mismatch time $\mu$ is small on average.
See also Fig. \ref{fig:Lemma_Explain}.
\begin{lem}
	\label{lem:tor0}
	Let Assumptions \ref{ass:system},
	\ref{ass:at_most_one_switch}, \ref{ass:quantization_near_origin}, and
	\ref{ass:subsystem_QA}
	hold, and let
	$L \geq 0$ satisfy \eqref{eq:L_inequality1}.
	If $\mu(t,0)$ achieves \eqref{eq:mu_inequality1}
	for all $t > 0$, then 
	there exists $T_{r} \geq 0$ such that 
	$x(T_{r}) \in \underline{\mathcal{E}}_P(r)$ for every
	$x(0) \in \Int(\overline{\mathcal{E}}_P(R))$
	and $\sigma(0) \in \mathcal{P}$, and furthermore
	$x (t) \in \Int (\overline{\mathcal{E}}_P(R))$ for all $t \in [0, T_{r}]$.
\end{lem}
\begin{pf}
	First we show that the trajectory $x(t)$ does not leave
	$\Int (\overline{\mathcal{E}}_P(R))$ without belonging 
	to $\underline{\mathcal{E}}_P(r)$. Namely, 
	there does not exist $T_{R} > 0$
	such that 
	\begin{gather}
		\label{eq:x(TR0)_outside}
		x(T_{R}) \in \partial \overline{\mathcal{E}}_P(R), \quad \text{and}\\
		%\end{equation}
		%and
		%\begin{equation}
		\label{eq:x(t)_0_TR0}
		x(t) \in 
		\Int (\overline{\mathcal{E}}_P(R))
		\setminus \underline{\mathcal{E}}_P(r)
		\qquad (0 \leq t < T_{R}).
	\end{gather}
	
	Assume, to reach a contradiction, \eqref{eq:x(TR0)_outside} and
	\eqref{eq:x(t)_0_TR0} hold for some $T_{R} > 0$.
	Recall that
	\begin{equation*}
		\lambda_{\min}(P) \|x\|^2 \leq V(x) = x^{\top} Px \leq 
		\lambda_{\max}(P) \|x\|^2 
	\end{equation*}
	for $x \in \mathbb{R}^{\sf n}$.
	%Since $x([t]^-) \in 
	%\overline{\mathcal{E}}_P(R)~\backslash~\underline{\mathcal{E}}_P(r)$
	%for all $t \in [0, T_{R})$ by \eqref{eq:x(t)_0_TR0}, 
	It follows from
	\eqref{eq:dotVp_bound} and \eqref{eq:dotVpq_bound} that
	\begin{align}
		\label{eq:VpVpq_bound_by_V}
		\begin{array}{c}
			\dot V_p(x(t),q_x(t)) \leq -C_PV(x(t)) \\[4pt]
			\dot V_{p,q}(x(t),q_x(t)) \leq D_P V(x(t)).
		\end{array}
		%\dot V_p(x(t),q_x(t)) \leq -C_PV(x(t)), \qquad
		%\dot V_{p,q}(x(t),q_x(t)) \leq D_P V(x(t)).
	\end{align}
	
	By 
	\eqref{eq:x(t)_0_TR0} and 
	\eqref{eq:VpVpq_bound_by_V}, a successive calculation at each switching time
	shows that
	\begin{align}
		\label{eq:VTR0_bound_by_V0}
		V(x(T_{R}))  
		&\leq 
		\exp \big(D_P \mu(T_{R},0)  - C_P(T_{R} - \mu(T_{R},0))\big) 
		V(x(0)).
	\end{align}
	Since \eqref{eq:mu_inequality1} gives
	\begin{align}
		%\hspace{-6pt}
		D_P \mu(t,0)-C_P(t - \mu(t,0))
		\leq
		\left(
		\left(
		C_P + 
		D_P
		\right) L - 
		C_P
		\right)t
		\label{eq:CD_bound0}
	\end{align}
	for all $t > 0$,
	it follows from \eqref{eq:L_inequality1} and 
	$x(0) \in \Int(\overline{\mathcal{E}}_P(R))$ that
	\begin{align*}
		%\label{eq:V(x(TR))_upper_bound}
		V(x(T_{R})) < V(x(0)) < R^2 \lambda_{\max} (P).
	\end{align*}
	However, \eqref{eq:x(TR0)_outside} shows that
	$
	%\label{eq:V(x(TR))_lower_bound}
	V(x(T_{R})) 
	%=\lim_{\varepsilon > 0, \varepsilon \to 0}V(x(T_{R} + \varepsilon))
	= R^2 \lambda_{\max} (P),
	$
	and we have a contradiction.
	
	Let us next prove that
	$x(T_{r}) \in \underline{\mathcal{E}}_P(r)$ for some
	$T_{r} \geq 0$.
	
	Suppose $x(t) \not\in \underline{\mathcal{E}}_P(r)$ for 
	all $t \geq 0$. Then since the discussion above shows that 
	$x(t) \in \Int (
	\overline{\mathcal{E}}_P(R))
	\setminus \underline{\mathcal{E}}_P(r)$
	for all $t \geq 0$, we obtain \eqref{eq:VTR0_bound_by_V0} with arbitrary
	$t \geq 0$ in place of $T_{R}$.
	Hence \eqref{eq:L_inequality1} and \eqref{eq:CD_bound0} show that
	$V(x(t)) \to 0$ as $t \to \infty$.
	However, this contradicts
	$x(t) \not\in \underline{\mathcal{E}}_P(r)$, i.e.,
	$V(x(t)) > r^2 \lambda_{\min}(P) > 0$. Thus there exists 
	$T_{r} \geq 0$ such that 
	$x(T_{r}) \in \underline{\mathcal{E}}_P(r)$.
\end{pf}

From the next result, we see that
the trajectory leaves $\underline{\mathcal{E}}_P(r)$ only if
a switch occurs between sampling times.
This is intuitively obvious because as mentioned in \cite{Ishii2004},
$\underline{\mathcal{E}}_P(r)$ is an invariant set
if a mode mismatch does not occur.
\begin{lem}
	\label{lem:cross_to_out}
	Let Assumptions \ref{ass:system},
	\ref{ass:at_most_one_switch}, \ref{ass:quantization_near_origin}, and
	\ref{ass:subsystem_QA} hold.
	If the trajectory $x(t)$ leaves
	$\underline{\mathcal{E}}_P(r)$ at $t = T_0$,
	more precisely, if
	there exists $\delta > 0$ such that
	\begin{equation}
		\label{eq:x(T_0)_def}
		x(T_0) \in \partial \underline{\mathcal{E}}_P(r),~
		x(T_0+\varepsilon) \not\in \underline{\mathcal{E}}_P(r)
		~~~~ (0 <  \varepsilon < \delta),
	\end{equation}
	then $\sigma(T_0) \not= \sigma([T_0]^-)$.
\end{lem}

\begin{pf}
	Assume, to get a contradiction,
	that $\sigma(T_0) = \sigma([T_0]^-)$.
	Suppose that 
	$\sigma(T) \not= \sigma([T]^-)$ for some $T > T_0$.
	Let $T_1$ be the smallest number of such $T$.
	Define an interval $I_\delta$ by 
	\[
	I_{\delta} := (0,\min \{\delta, T_1-T_0\}).
	\]
	If there does not exist $T > T_0$ with $\sigma(T) \not= \sigma([T]^-)$,
	then we define $I_\delta$ by $I_\delta := (0,\delta)$.
	Since $V(x(t))$ is differentiable at all $t \geq 0$ except for sampling times and
	switching times,
	there is no loss of generality in assuming that 
	$V(x(t))$ is differentiable in $I_{\delta}$.
	Since $\sigma(T_0 + \varepsilon) = \sigma([T_0 + \varepsilon]^-) = \sigma([T_0]^-)$
	for all $\varepsilon \in I_{\delta}$,
	it follows from \eqref{eq:dotVp_bound} that
	\begin{equation*}
		%\label{eq:dotV_T0+eps}
		\dot V(x(T_0+\varepsilon)) \leq - C \|x((T_0+\varepsilon))\|^2 \leq 0
		\qquad (\varepsilon \in I_{\delta}).
	\end{equation*}
	However, 
	\eqref{eq:x(T_0)_def} gives
	%$x(T_0 + \varepsilon) \not\in \underline{\mathcal{E}}_P(r)$,
	\begin{equation*}
		%\label{eq:V_T0+eps}
		V(x(T_0+\varepsilon)) > r^2 \lambda_{\min}(P) = V(x(T_0))
		\qquad (\varepsilon \in I_{\delta}).
	\end{equation*}
	Since $V(x(t))$ is continuous, we have a contradiction
	by the mean value theorem. Thus $\sigma(T_0) \not= \sigma([T_0]^-)$.
\end{pf}

Lemma \ref{lem:r_to_ar} below shows that 
the trajectory stays in a slightly larger ellipsoid than 
$\underline{\mathcal{E}}_P(r)$
after the trajectory enters into $\underline{\mathcal{E}}_P(r)$; 
see Fig. \ref{fig:Lemma_Explain}.  
\begin{lem}
	\label{lem:r_to_ar}
	Let Assumptions \ref{ass:system},
	\ref{ass:at_most_one_switch}, \ref{ass:quantization_near_origin}, and
	\ref{ass:subsystem_QA} hold.
	Suppose that $T_0 \geq 0$ is a time at which
	$x(t)$ leaves $\underline{\mathcal{E}}_P(r)$.
	Let $\kappa > 1$ satisfy \eqref{eq:a_cond}
	and define $f(\kappa )$ by \eqref{eq:ba_bound}.
	Pick $L \geq 0$ with \eqref{eq:L_inequality1}.
	If $\mu(t,T_0)$ satisfies
	\eqref{eq:mu_b_L_condition}
	for all $t > T_0$, then for every $\sigma(T_0) \in \mathcal{P}$,
	there exists $T_{1} > T_0$ such that 
	$x(T_1) \in \underline{\mathcal{E}}_P(r)$, and furthermore
	$x (t) \in \Int(\underline{\mathcal{E}}_P(\kappa r))$ 
	for all $t \in [T_0, T_1]$.
\end{lem}
%The proof follows in a manner similar to that of Lemma \ref{lem:tor0}, 
%so we omit it.
\begin{pf}
	By \eqref{eq:mu_b_L_condition},
	$V(x(t))$ satisfies
	\begin{align}
		&V(x(t)) \leq 
		\exp\big( 
		\left(
		\left(
		C_P+ D_P \right)L - 
		C_P
		\right) (t - T_0)
		\big) \cdot  \exp\big(
		\left(C_P  + D_P\right)f(\kappa)
		\big) V(x(T_0) ) \label{eq:Vt_VT0}
	\end{align}
	if $t > T_0$ satisfies
	$x(t')\in
	\overline{\mathcal{E}}_P(R)
	\setminus\underline{\mathcal{E}}_P(r)$
	for all $t' \in (T_0, t]$.
	On the other hand,
	since $x(T_0) \in \partial \underline{\mathcal{E}}_P(r) $,
	it follows from \eqref{eq:ba_bound} that
	\begin{align}
		\label{eq:Lyapunov_bound_ar0}
		\exp\big(
		\left(C_P + D_P\right)f(\kappa)
		\big) V(x(T_0) ) 
		= \kappa^2 r^2 \lambda_{\min}(P).
	\end{align}
	%Since \eqref{eq:a_cond} holds if and only if
	%$
	%\underline{\mathcal{E}}_P(ar) 
	%%= 
	%%\{
	%%x \in \mathbb{R}^{\sf n}:~V(x) \leq a^2e^{\bar D_P}r^2\lambda_{\min}(P)
	%%\}
	%\subset \Int (\overline{\mathcal{E}}_P(R)),
	%$
	In conjunction with \eqref{eq:a_cond},
	this leads to 
	\[
	\exp\big(
	\left(C_P + D_P\right)f(\kappa)
	\big) V(x(T_0) )
	< R_2 \lambda_{\max}(P).
	\] 
	Hence we have
	$x(T_1) \in \underline{\mathcal{E}}_P(r)$ for some
	$T_1 > T_0$
	from \eqref{eq:L_inequality1} and 
	\eqref{eq:Vt_VT0}
	as in the proof of Lemma \ref{lem:tor0}. 
	Substituting \eqref{eq:Lyapunov_bound_ar0} into \eqref{eq:Vt_VT0},
	we also obtain 
	$V(x(t)) < \kappa^2 r^2\lambda_{\min}(P)$ for $t \geq T_0$. 
	Thus
	$x(t) \in \Int (\underline{\mathcal{E}}_P(\kappa r))$ 
	for $t \in [T_0, T_1]$.
\end{pf}

Finally, we prove Theorem \ref{thm:main_thm} by using
Lemmas \ref{lem:tor0}, \ref{lem:cross_to_out}, and \ref{lem:r_to_ar}:

\begin{spf}
	Lemma \ref{lem:tor0} shows that if \eqref{eq:mu_inequality1} holds
	for all $t > 0$, then
	$x(T_{r}) \in \underline{\mathcal{E}}_P(r)$ for some $T_r > 0$ and
	$x (t) \in \Int (\overline{\mathcal{E}}_P(R))$ for all $t \in [0, T_r]$.
	%If we let $\tau_1, \tau_2, \dots$ be times at which $x(t)$ leaves 
	%$\underline{\mathcal{E}}_P(r)$, then $\tau_{k+2} - \tau_{k} > T_s$ by
	%the switching condition in Assumption \ref{ass:system}, and hence
	%$\tau_k \to \infty$ as $k \to \infty$.
	%%, and hence
	%%the trajectory does not chatter on $\partial \underline{\mathcal{E}}_P(r)$.
	%Using Lemmas \ref{lem:cross_to_out} and \ref{lem:r_to_ar} 
	%at each $\tau_1, \tau_2, \dots$,
	%we have that 
	%if \eqref{eq:mu_b_L_condition} holds
	%for all $t > T_0$ with 
	%$\sigma(T_0) \not= \sigma([T_0]^-)$, then
	%$x(t) \in \Int(\underline{\mathcal{E}}_P(\kappa r))$ 
	%for all $t \geq T_{r}$. 
	Let $\tau_1, \tau_2, \dots$ be the instants at which $x(t)$ leaves 
	$\underline{\mathcal{E}}_P(r)$.
	Using Lemmas \ref{lem:cross_to_out} and \ref{lem:r_to_ar}
	at each $\tau_1, \tau_2, \dots$,
	we have that if for each $T_0 \geq 0$ with 
	$\sigma(T_0) \not= \sigma([T_0]^-)$,
	\eqref{eq:mu_b_L_condition} holds
	for every $t > T_0$, then
	there exists $\hat{\tau}_k \in (\tau_k, \tau_{k+1}]$ such that
	$x(\hat{\tau}_k) \in \underline{\mathcal{E}}_P(r)$ 
	and $x(t) \in \Int(\underline{\mathcal{E}}_P(\kappa r))$
	for all
	$t \in [\tau_k, \tau_{k+1}]$.
	Hence
	if $\{\tau_k\}$ has only finitely many elements,
	then the stability is achieved.
	On the other hand, if we have infinitely many $\tau_k$, then
	$\tau_k \to \infty$ as $k \to \infty$, because
	$\tau_{k+2} - \tau_{k} > T_s$ by
	the switching condition in Assumption
	\ref{ass:at_most_one_switch}.
	Thus $x(t) \in \Int(\underline{\mathcal{E}}_P(\kappa r))$ 
	for all $t \geq T_{r}$.
	This completes the proof.
\end{spf}

%Referring to 
%Lemmas \ref{lem:tor0}, \ref{lem:cross_to_out}, and \ref{lem:r_to_ar},
%we immediately derive the following result:

\section{Reduction to a Dwell-Time Condition}
In the preceding section, we have derived a sufficient condition 
on the total mismatch time $\mu$ for the stability of
the quantized sampled-data systems with multiple modes.
However, it may be difficult to check whether $\mu$ satisfies
\eqref{eq:mu_inequality1} and \eqref{eq:mu_b_L_condition}.
In this section, we will show that these conditions \eqref{eq:mu_inequality1} and \eqref{eq:mu_b_L_condition}
can be achieved for
switching signals with a certain dwell-time property.

%The proofs of theorems in this section are omitted for space reason.

To proceed, we recall the definition of dwell time:
We call $\sigma$ 
a \textit{switching signal with dwell time $T_d$}
if the switching signal $\sigma$ has an interval between
consecutive discontinuities no smaller than $T_d > 0$ and further if $\sigma$ 
has no discontinuities in $[0, T_d)$.

%To proceed, we recall the precise definition of a \textit{switching signal with the dwell time $T_d$}.
%Let $\rho:~[0,\infty) \to [0,\infty)$ satisfy $\rho(0)=0$ and $\dot \rho(t) = 1$ for
%$t \geq 0$. The switching signal $\sigma$ has the dwell time $T_d$ if
%\begin{align*}
%&\rho(t) < T_d  \quad \Rightarrow \quad
%\sigma(t) = \sigma(t-\tau) \quad ( 0 \leq \tau \leq \rho(t)),\\
%&\sigma(t) \not= \lim_{\epsilon >0,\epsilon\to 0}\sigma(t-\epsilon)
%\quad \Rightarrow \quad \rho(t) = 0.
%\end{align*}
%Note that a switching signal with a dwell time
%has no discontinuities in the first dwell-time period.

The following proposition gives 
an upper bound of the total mismatch time
for switching signals with dwell time.
\begin{prop}
	\label{thm:mu_upper_bound}
	Fix $n \in \mathbb{N}$.
	For every switching signal
	$\sigma$ with dwell time $nT_s$, $\mu$ in \eqref{eq:mu_def_same}
	satisfies
	\begin{equation}
		\label{eq:mu_0T} 
		\mu(t,0) < \frac{t}{n} \qquad (t > 0).
	\end{equation}
	Furthermore, if $\sigma(T_0) \not= \sigma([T_0]^-)$, then 
	\begin{equation} 
		\label{eq:mu_T0T}
		\mu(t,T_0) < T_s + \frac{t-T_0}{n}
		\qquad (t > T_0).
	\end{equation}
\end{prop}

\begin{pf}
	The proof includes a lengthy but routine calculation;
	see Appendix A.1.
\end{pf}

Theorem \ref{thm:main_thm} and 
Proposition \ref{thm:mu_upper_bound} can be 
combined in the following way:
\begin{thm}
	\label{thm:dwell_sample_switching}
	Let Assumptions \ref{ass:system},
	\ref{ass:at_most_one_switch}, \ref{ass:quantization_near_origin}, and
	\ref{ass:subsystem_QA} hold.
	Let $n \in \mathbb{N}$ satisfy $n \geq 1+D_P/C_P$.
	Define 
	\begin{equation}
		\label{eq:kappa_dwell}
		\kappa := \exp 
		\left(
		\frac{T_s(C_P + D_P)}{2}
		\right),
	\end{equation}
	and suppose that $\kappa$ satisfies \eqref{eq:a_cond}.
	%and suppose that
	%\begin{align}
	%\label{eq:a_requirement}
	%a \leq \frac{R}{r}\sqrt{\frac{\lambda_{\max}(P)}{\lambda_{\min}(P)}}.
	%\end{align}
	%If $x(0) \in \Int(\overline{\mathcal{E}}_P(R))$
	%and if the dwell time of $\sigma$ is $nT_s$, then
	%every trajectory $x(t)$ of the switched system \eqref{eq:SLS} 
	%with \eqref{eq:control_input} satisfies
	%$x(t) \in \Int(\overline{\mathcal{E}}_P(R))$ for all $t \geq 0$, and furthermore
	%there exists $T_{r} \geq 0$ such that 
	%$x(t) \in \Int(\underline{\mathcal{E}}_P(\kappa r))$ for all $t \geq T_{r}$.
	If the dwell time of $\sigma$ is $nT_s$, then
	there exists $T_{r} \geq 0$ such that 
	for every $x(0) \in \Int(\overline{\mathcal{E}}_P(R))$ and
	$\sigma(0) \in \mathcal{P}$,
	$x(t) \in \Int(\underline{\mathcal{E}}_P(\kappa r))$ 
	for all $t \geq T_{r}$.
	Furthermore, $x(t) \in \Int(\overline{\mathcal{E}}_P(R))$ for all $t \geq 0$.
\end{thm}
\begin{pf}
	If $n$ and $\kappa$ are defined as above, 
	Proposition \ref{thm:mu_upper_bound} shows that
	$\mu$ satisfies
	\eqref{eq:mu_inequality1} and
	\eqref{eq:mu_b_L_condition} for every 
	switching signal $\sigma$ with dwell time $nT_s$.
	%Also, $\underline{\mathcal{E}}_P(ar) \subset 
	%\overline{\mathcal{E}}_P(R)$ follows from \eqref{eq:a_requirement}.
	Hence the conclusion of Theorem \ref{thm:main_thm} holds.
\end{pf}

The next result implies that the upper bounds obtained in 
Proposition \ref{thm:mu_upper_bound} are close to the supremum 
over all switching signals with dwell time $nT_s$
if the sampling period $T_s$ is sufficiently small.
\begin{prop}
	\label{thm:mu_lower_bound}
	Fix $\varepsilon > 0$ and $n \in \mathbb{N}$.
	For any $T \geq 0$,
	there exist a switching signal $\sigma$ with dwell time $nT_s$ and $t \geq T$ such that
	\begin{align*}
		%\label{eq:lowerbound1}
		\mu(t,0) \geq \frac{t}{n} - \left( \frac{T_s}{n} + \varepsilon \right).
	\end{align*}
	Furthermore, for any $T \geq 0$,
	there exist a switching signal $\sigma$ 
	with dwell time $nT_s$, $T_0 \geq 0$ with
	$\sigma(T_0) \not= \sigma([T_0]^-)$, and
	$t \geq T_0 + T$ such that 
	\begin{align}
		\label{eq:lowerbound2}
		\mu(t, T_0) \geq T_s + \frac{t-T_0}{n} - 
		\left( \frac{T_s}{n} + \varepsilon \right).
	\end{align}
\end{prop}
\begin{pf}
	This is again a routine calculation; see Appendix A.2.
\end{pf}

The next result is the case $n=1$ in 
Proposition \ref{thm:mu_lower_bound}.
\begin{cor}
	\label{cor:small_dwelltime}
	There exist a switching signal 
	$\sigma$
	with dwell time $T_s$ such
	that $\mu(t,0) \approx t$ for sufficiently large $t>0$.
\end{cor}
This corollary shows that, not surprisingly, 
if the dwell time does not exceed the sampling period, then
the information on switching signals is not so useful for the 
stabilization of the 
sampled-data switched system.

\section{Numerical Example}
Consider the switched system with the following two modes:
\begin{gather*}
	A_1 = 
	\frac{1}{6}
	\begin{bmatrix}
		1 & -2\\ -3 & 2
	\end{bmatrix}, \quad
	B_1 = 
	\frac{1}{6}
	\begin{bmatrix}
		-4 \\ 3
	\end{bmatrix}\\
	A_2 = 
	\begin{bmatrix}
		1 & -5\\ 1 & 2
	\end{bmatrix}, \quad
	B_2 = 
	\begin{bmatrix}
		1 \\ -1
	\end{bmatrix}.
\end{gather*}
The state feedback gains $K_1$ and $K_2$ are given by
\begin{equation}
	\label{eq:state_feedback_gain_Ex}
	K_1 = \begin{bmatrix}
		1.38 &  -1.86
	\end{bmatrix},\quad
	K_2 = \begin{bmatrix}
		-2.80 &  3.77
	\end{bmatrix}.
\end{equation}
We computed the above regulator gains by minimizing the cost
\begin{align*}
	\int^{\infty}_{0} \big( x(t)^{\top}x(t) + u(t)^2 \big) dt.
\end{align*}
Note that both $A_1 + B_1K_2$ and $A_2 + B_2K_1$ are not Hurwitz:
$A_1 + B_1K_2$ has one unstable eigenvalue $4.4538$ and
$A_2 +B_2K_1$ has two unstable eigenvalues $1.4091$ and
$4.7750$.

The sampling period $T_s$ was given by $T_s = 0.025$, 
and we used the following logarithm quantizer:
Let the state $x$ be 
%$ x = 
%\begin{bmatrix}
%x_1 & x_2
%\end{bmatrix}^{\top}$.
$x = [x_1~~x_2]^{\top}$.
For a nonnegative integer $n$, 
the quantized state 
%$Q(x) =
%\begin{bmatrix}
%Q_1 & Q_2
%\end{bmatrix}^{\top}$
$Q(x) = [Q_1(x_1)~~Q_2(x_2)]^{\top}$
is defined by
\begin{equation*}
	Q_i(x_i) :=
	\begin{cases}
		\frac{-\xi_0 (\eta^n + \eta^{n+1})}{2}
		&  (-\xi_0 \eta^{n+1} \leq x_i < -\xi_0\eta^n)\\ 
		0
		& (-\xi_0 \leq x_i \leq \xi_0) \\
		\frac{\xi_0 (\eta^n + \eta^{n+1})}{2}
		& (\xi_0 \eta^n < x \leq \xi_0\eta^{n+1}),
	\end{cases}
\end{equation*}
where $\xi_0 = 0.08$ and $\eta= 1.2$.

Set $C = 1$, $R=68.6$, and $r=0.175$ in Assumption~\ref{ass:subsystem_QA}.
Algorithm \ref{alg:commonLyapunov} of Appendix B 
gave 
the positive definite matrix $P$ in Assumption~\ref{ass:subsystem_QA}
by
\begin{equation*}
	%\label{eq:P_value}
	P = 
	\begin{bmatrix}
		2.9171 & 0.3489 \\ 
		0.3489 & 3.6256
	\end{bmatrix}.
\end{equation*}
In the randomized algorithm, 
we used $10^7$ samples in state for each run,
and five samples in time for each sampled state.
We stopped the algorithm when there was no update for an entire run.

Since we obtain $D = 55.15$ in \eqref{eq:dotVpq_bound} from the data above,
the resulting $n$ and $\kappa$ in Theorem \ref{thm:dwell_sample_switching} are
$n=76$ and $\kappa=1.2864$.

\begin{figure}
	\centering
	\subcaptionbox{Region $[-80, 80] \times [-75, 75]$ \label{fig:LQ11fig1}}
	[.49\linewidth]
	{\includegraphics[width = 8.5cm,clip]{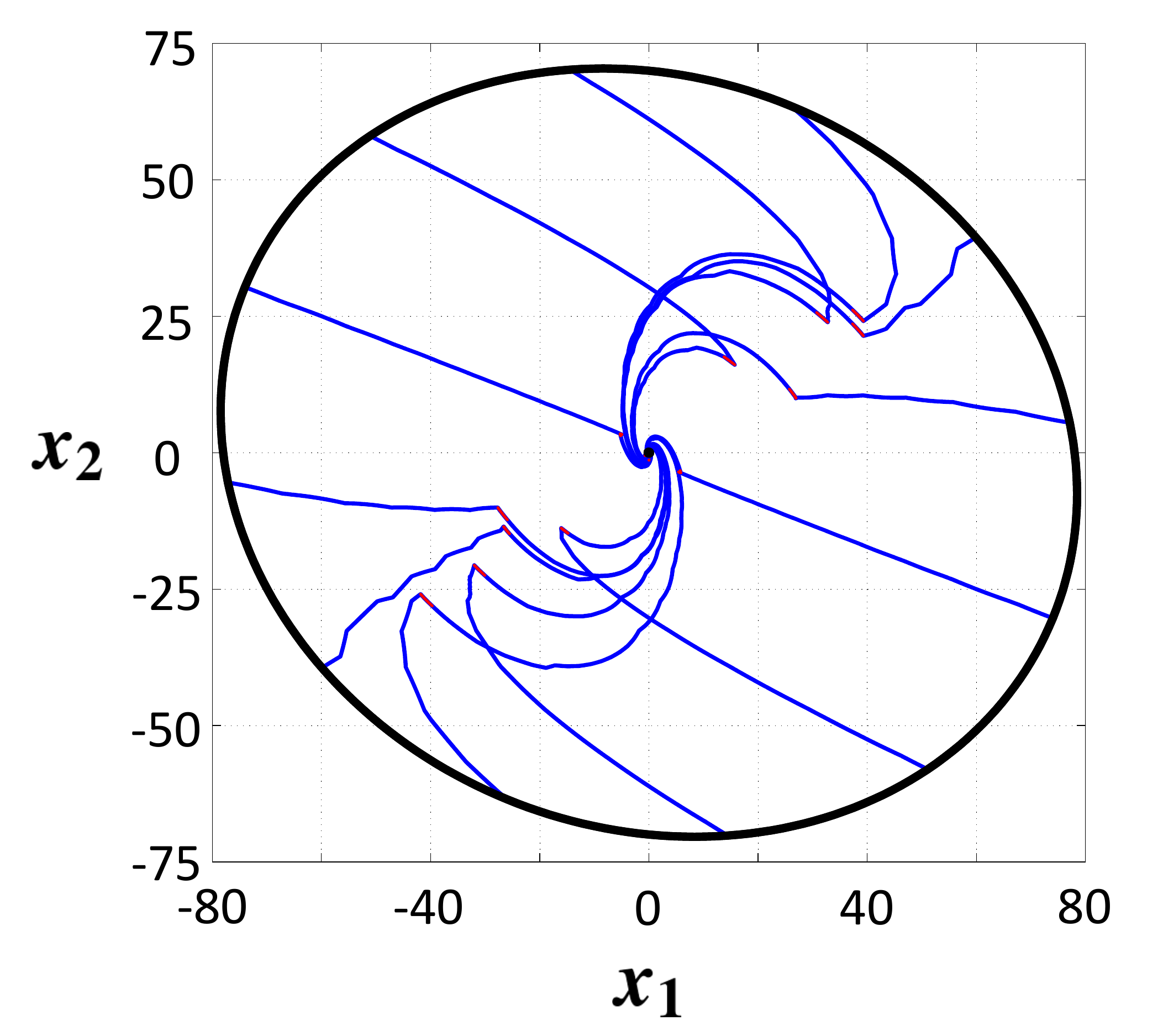}} 
	\subcaptionbox{Region $[-0.75, 0.75] \times [-1.0, 1.0]$ \label{fig:LQ11fig3}}
	[.5\linewidth]
	{\includegraphics[width = 8.2cm,clip]{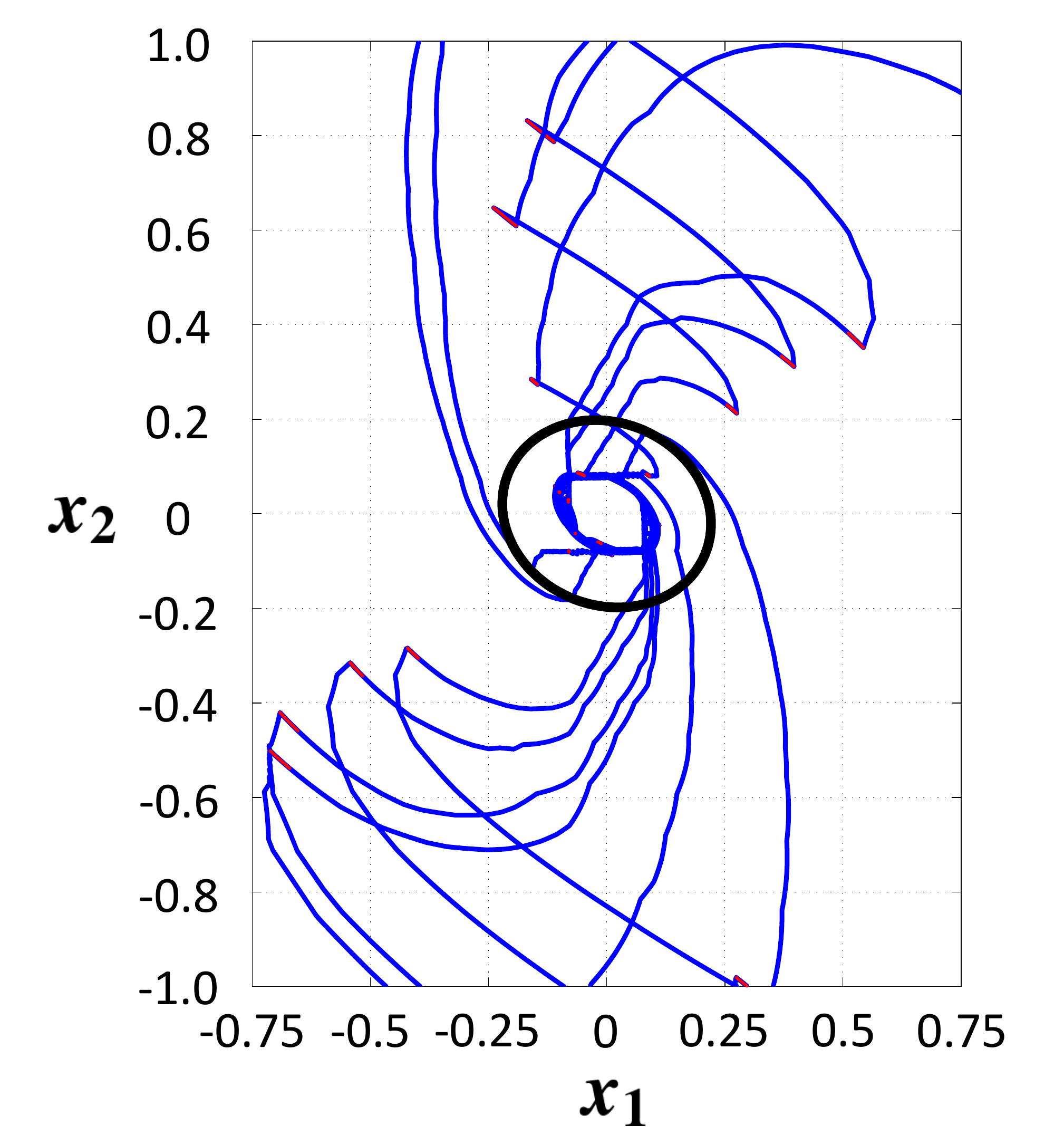}}
	\caption{The trajectories $x$ with 
		$\sigma(0)=1$ 
		\label{fig:LQ11fig}}
\end{figure}

A time response ($0 \leq t \leq 20$) was calculated for 
$\sigma(0)=1$ and some initial states on $\partial \overline{\mathcal{E}}_P(R-\epsilon)$
with $\epsilon = 0.001$.
Fig.~\ref{fig:LQ11fig} depicts the state trajectories $x$ 
of the switched system~\eqref{eq:SLS} with dwell time $76T_s = 1.9$.
After an interval of length $76T_s$ with no switches,
a switch of the plant mode occurs with probability 0.05 per sampling interval
and the distribution is uniform in a sampling interval.
The blue line indicates that the feedback gain designed for the active subsystem
was used, i.e.,
\begin{equation*}
	(A_{\sigma(t)},B_{\sigma(t)},K_{\sigma([t]^-)})
	=
	(A_{1},B_{1},K_{1})
	~\text{or}~
	(A_{2},B_{2},K_{2}).
\end{equation*}
The red line shows that a switch led to the mismatch of the modes between
the plant and the feedback gain, i.e.,
\begin{equation*}
	(A_{\sigma(t)},B_{\sigma(t)},K_{\sigma([t]^-)})
	=
	(A_{1},B_{1},K_{2})
	~\text{or}~
	(A_{2},B_{2},K_{1}).
\end{equation*}
The black lines in Fig.~\ref{fig:LQ11fig}
represent the ellipsoid of initial conditions $\overline{\mathcal{E}}_P(R)$ 
and the attractor set
$\underline{\mathcal{E}}_P(\kappa r)$, respectively.

Here we see two conservative results: the dwell time $76T_s$
and the attractor set $\underline{\mathcal{E}}_P(\kappa r)$ 
in Fig.~\ref{fig:LQ11fig3}.
Since we evaluate the increasing and decreasing rates of the Lyapunov function 
only by \eqref{eq:dotVpq_bound} and \eqref{eq:dotVp_bound},
the switching condition for stability becomes conservative.
In particular, we need to refine the upper bound \eqref{eq:dotVpq_bound}
in the mismatch case, which has been obtained by assuming 
that we have the worst-case trajectory
whenever a mode mismatch occurs. If 
we know where switching happens as for piecewise affine systems, 
then the upper bound \eqref{eq:dotVpq_bound}
can be improved.

As regards the attractor set $\underline{\mathcal{E}}_P(\kappa r)$,
the trajectories in Fig.~\ref{fig:LQ11fig3} stayed in
a smaller neighborhood of the origin. 
The conservative result is also due to 
the upper bound \eqref{eq:dotVpq_bound}; 
see \eqref{eq:kappa_dwell}.
Another reason is 
the nonlinearity of static quantizers
and this conservatism is
observed for systems with a single mode as well \cite{Ishii2002Book, Ishii2004, 
	Haimovich2007}.
Construction of polynomial Lyapunov functions may allow us to obtain
less conservative bounds.

If we use multiple Lyapunov functions together with
an average dwell-time property,
instead
of a common Lyapunov function, then 
the above conservatism can be reduced.
On the other hand,
the authors of \cite{Haimovich2010} have proposed 
the calculation method of an ultimate bound and an invariant set
for continuous-time switched systems with disturbances.
If one can generalize this method to 
sampled-data switched systems with a static quantizer, then
another insight into the state trajectory near the origin will be obtained.
Details, however, are more involved, 
so these extensions are subjects for future research.

%\begin{figure}[b]
%\centering
%\subcaptionbox{Region $[-70, 10] \times [-10, 60]$ \label{fig:LQ11fig1}}
%{\includegraphics[width = 6.5cm,clip]{LQ11new_verC_fig1.pdf}}
%\vspace{15pt}
%
%\subcaptionbox{Region $[-0.5, 0.3] \times [-0.8, 0.3]$ \label{fig:LQ11fig3}}
%{\includegraphics[width = 5cm,clip]{LQ11new_verC_fig2.pdf}}
%%\vspace{15pt}
%%
%%\subcaptionbox{Region $[-0.25, 0.25] \times [-0.25, 0.25]$ \label{fig:LQ11fig4}}
%%{\includegraphics[width = 6.5cm,bb= 30 0 650 540,clip]{LQ11new_verA_4.pdf}}
%\caption{The trajectory $x$ with $x(0) = [-60~~50]^{\top}$ and $\sigma(0)=1$
%\label{fig:LQ11fig}}
%\end{figure}

\section{Concluding Remarks}
For sampled-data switched systems 
with static quantizers,
we have developed a
stability analysis by using a common Lyapunov function
computed efficiently from a randomized algorithm.
We have derived a switching condition on the total mismatch time,
and have found a neighborhood of the origin into which all trajectories
fall whenever the initial state is within a known bound. 
Moreover,
the condition on the total mismatch time has been
reduced to a dwell-time condition.
Future work will focus on improving the upper bound on the growth rate 
of the Lyapunov function in the mismatched case, and
analyzing the stability by
multiple Lyapunov functions and an average dwell-time property.
%which can reduce conservativeness in our proposed method.
%Stabilization by quantized output feedback is currently being examined.

\appendix
\section{Bound on Total Mismatch Time}    % Each appendix must have a short title.
\subsection{Proof of Proposition \ref{thm:mu_upper_bound}} 
Let us first prove \eqref{eq:mu_0T}.
It is clear that $\mu = 0$ if $\sigma$ has no discontinuities in the interval $(0,t)$.
Let $t_1,\dots,t_m$ be the switching times in $(0,t)$. We have
%The mismatch time in 
%the interval $[ [t_k]^-, [t_{k+1}]^- )$ is $[t_k]^- + T_s - t_k$
\begin{equation*}
	\mu([t_{k+1}]^-, [t_k]^-) = 
	\begin{cases}
		[t_k]^- + T_s - t_k & \text{if $t_k \not= [t_k]^-$} \\
		0 & \text{otherwise}
	\end{cases}
\end{equation*}
for $k=1,\dots,m-1$, 
and 
\begin{equation*}
	\mu(t, [t_m]^-) = 
	\begin{cases}
		[t_m]^- + T_s - t_m  & \text{if $t_m \not= [t_m]^-$ and
			$[t_m]^- +T_s < t$} \\
		t- t_m & \text{if $t_m \not= [t_m]^-$ and
			$[t_m]^- +T_s \geq t$} \\
		0 & \text{otherwise}
	\end{cases}
\end{equation*}
%that in the 
%interval $[ [t_m]^-, t )$ is 
%$\min \{ [t_m]^- + T_s - t_m,~t-t_m\}$.
Since $t \geq mnT_s$, we obtain
\begin{equation*}
	\mu(t,0) \leq \sum_{k=1}^{m} ([t_k]^-+T_s - t_k)
	< mT_s \leq \frac{1}{n}t.
\end{equation*}
Hence \eqref{eq:mu_0T} holds.

Next we show \eqref{eq:mu_T0T}.
Since $\sigma(T_0) \not= \sigma([T_0]^-)$ and since
the dwell time is $nT_s \geq T_s$, it follows that
$\sigma$ has precisely one discontinuity in the interval $([T_0]^-, T_0]$.
Let us denote the switching time by $t_0$.

Suppose that no switches occur in the interval $(T_0, t)$.
Since only the interval $[T_0, [T_0]^- + T_s)$ has a mode mismatch,
it follows that
\begin{equation*}
	\mu(t, T_0) \leq [T_0]^- +T_s - T_0 < T_s,
\end{equation*}
and hence \eqref{eq:mu_T0T} holds.

Suppose that $m$ switches occur in the interval $(T_0, t)$, and let
$t_1,\dots,t_m$ be the switching times.
Define $\xi_k$ by
\begin{equation}
	\label{eq:xi_def}
	\xi_k := (t_{k+1} - t_k ) - nT_s
\end{equation}
for $k=0,\dots,m-1$.
The dwell-time assumption implies that $\xi_k \geq 0$.
We also have
\begin{align}
	t - T_0 &= 
	(t - t_m) + \sum_{k=0}^{m-1}(t_{k+1} - t_{k}) - (T_0 - t_0) \notag \\
	&=
	(t - t_m) + \sum_{k=0}^{m-1}(\xi_{k} + nT_s) - (T_0 - t_0) \notag\\
	&=
	mnT_s + (t-t_m) + \sum_{k=0}^{m-1}\xi_{k} - (T_0 -t_0).
	\label{eq:t-T_0}
\end{align}
We split the argument into two cases:
\begin{equation}
	\label{eq:case1}
	(t-t_m) + \sum_{k=0}^{m-1}\xi_k \geq T_0 -t_0
\end{equation}
and
\begin{equation}
	\label{eq:case2}
	(t-t_m) + \sum_{k=0}^{m-1}\xi_k < T_0 -t_0.
\end{equation}

First we study the case \eqref{eq:case1}, where
some switching intervals are sufficiently larger than $nT_s$.
Combining \eqref{eq:case1} with \eqref{eq:t-T_0}, 
we obtain $t-T_0 \geq mnT_s$, and hence
\begin{align*}
	\mu(t, T_0) &\leq  ([T_0]^-+T_s - T_0)
	+ \sum_{k=1}^{m} ([t_k]^-+T_s - t_k) \\
	&< (m+1)T_s \leq T_s + \frac{1}{n}(t-T_0),
\end{align*}
which is a desired inequality \eqref{eq:mu_T0T}.

Let us next consider the case \eqref{eq:case2}, where
every switching interval is smaller than $nT_s$.
Since 
\begin{align*}
	\mu(t, T_0) 
	= \mu([t_1]^-, T_0) &+ \sum_{k=1}^{m-1} \mu([t_{k+1}]^-, [t_k]^-)
	+ \mu(t, [t_m]^-)
\end{align*}
and since $\mu([t_1]^-, T_0) = \mu([T_0]^- + T_s,  T_0) \leq [T_0]^- + T_s - T_0$,
it is enough to obtain upper bounds on $\mu([t_{k+1}]^-, [t_k]^-)$
and $\mu(t, [t_m]^-)$.

We first derive
\begin{equation}
	\mu([t_{k+1}]^-, [t_k]^-)
	\leq
	[t_0]^- +T_s - t_0
	\label{eq:t_k-t_k+1_bound}
\end{equation}
for $k=1,\dots,m-1$ as follows.
Since $\sum_{k = 0}^{m-1}\xi_{k} < T_0$ by \eqref{eq:case2},
each switching time $t_k$ ($k=1,\dots,m$) satisfies
\begin{align*}
	t_k-t_0 
	&= (t_k - t_{k-1}) + \dots + (t_1 - t_0)  \\
	%&= knT_s + \sum_{\ell = 0}^{k-1}\xi_{\ell }
	&= \sum_{\ell = 0}^{k-1} (\xi_{\ell} + nT_s) \\
	&\leq  \sum_{\ell = 0}^{m-1}\xi_{\ell} + knT_s \\
	&< T_0 - t_0 + knT_s.
\end{align*}
In conjunction with the assumption on the dwell time, 
this leads to
\begin{equation}
	\label{eq:t_k_bound}
	t_0+knT_s \leq t_k < T_0+knT_s
\end{equation}
for every $k=1,\dots,m$. 
Since 
\begin{equation}
	\label{eq:t0-T0}
	[t_0]^- = [T_0]^- < t_0 \leq T_0 < [T_0]^- + T_s, 
\end{equation}
\eqref{eq:t_k_bound} shows that $[t_k]^- = [t_0]^- + knT_s$, and hence
\begin{equation*}
	t_0 + knT_s \leq t_k < [t_k]^- + T_s = [t_0]^- + knT_s + T_s,
\end{equation*}
which gives $[t_k]^- + T_s - t_k \leq [t_0]^- +T_s - t_0$.
We therefore have
\begin{align}
	\mu([t_{k+1}]^-, [t_k]^-) 
	&=
	\mu([t_k]^-+T_s, [t_k]^-) \notag \\
	&\leq
	[t_k]^- + T_s - t_k \notag \\
	&\leq
	[t_0]^- +T_s - t_0. \notag
\end{align}
Thus we obtain \eqref{eq:t_k-t_k+1_bound}.

Similarly, we can obtain
\begin{equation}
	\label{eq:t-t_m_bound}
	\mu(t, [t_m]^-) < T_0 - t_0.
\end{equation}
In fact, \eqref{eq:case2} and \eqref{eq:xi_def} give
\begin{equation*}
	t < (T_0 - t_0) + t_m -\sum_{k=0}^{m-1} \xi_{k}
	= T_0 + mnT_s.
\end{equation*}
If we combine this with
$t > t_m$ and \eqref{eq:t_k_bound},
we see that
\begin{equation*}
	t_0 + mnT_s 
	\leq t_m < t < T_0 + mnT_s,
\end{equation*}
which implies that
\begin{equation*}
	\mu(t, [t_m]^-) \leq t - t_m< T_0 - t_0.
\end{equation*}
We therefore have \eqref{eq:t-t_m_bound}.

Since $t- t_m > 0$ and $\xi_k\geq 0$,
it follows from \eqref{eq:t-T_0} that $m$ satisfies
$t - t_0 > mnT_s$, i.e.,
\begin{align}
	\label{eq:mbound}
	m < \frac{t-t_0}{nT_s}.
\end{align}
By \eqref{eq:t_k-t_k+1_bound},
\eqref{eq:t-t_m_bound}, and \eqref{eq:mbound}, we have
\begin{align}
	\mu(t, T_0) 
	&< ([T_0]^-+T_s - T_0)  + (m-1)([t_0]^-+T_s - t_0)+ (T_0 - t_0) \notag \\
	&< \frac{t-t_0}{n}\frac{[t_0]^-+T_s - t_0}{T_s} \notag \\
	&< \frac{t-[t_0]^-}{n}.
	\label{eq:mu(T_0,t)_case2_bound}
\end{align}
Moreover, \eqref{eq:t0-T0} gives
\begin{align*}
	T_s + \frac{t-T_0}{n}- \frac{t-[t_0]^-}{n}
	&= T_s - \frac{T_0 - [t_0]^-}{n} \\
	&> T_s - \frac{T_s}{n} \geq 0.
\end{align*}
Hence 
\eqref{eq:mu_T0T} follows from \eqref{eq:mu(T_0,t)_case2_bound}.

\subsection{Proof of Proposition \ref{thm:mu_lower_bound}}
Fix $T \geq 0$ and 
suppose that $m \in \mathbb{N}$ satisfies $mnT_s \geq T$.

To prove the first assertion of the theorem, let a switching signal
$\sigma$ have discontinuities at $knT_s + \varepsilon/m$ $(k=1,\dots,m)$.
If we define
$t := mnT_s+T_s$, then $t \geq T$ and we obtain
\begin{equation*}
	\mu(t, 0) = m\left(T_s - \frac{\varepsilon}{m}\right) = mT_s -\varepsilon =
	\frac{t}{n} - \left( \frac{T_s}{n} + \varepsilon \right).
\end{equation*}

To prove the second assertion, let $T_0 - [T_0]^- = \varepsilon/(2m+1)$ and let
$\sigma$ have a switch at
\begin{equation*}
	T_0+ knT_s + \frac{\varepsilon}{2(m+1)}
	= [T_0]^- + knT_s + \frac{\varepsilon}{m+1}.
\end{equation*}
for each $k=1,\dots,m$.
If we set $t := T_0+mnT_s+T_s$, then $t \geq T_0 +T$ and we have
\begin{align*}
	\mu(t,T_0) 
	&= \left(T_s - \frac{\varepsilon}{2(m+1)}\right) + 
	m\left(T_s - \frac{\varepsilon}{m+1} \right) \\
	&\geq (m+1)T_s - \varepsilon \\
	&= T_s + \frac{t-T_0}{n}
	- \left( \frac{T_s}{n} + \varepsilon\right),
\end{align*}
which is the desired inequality \eqref{eq:lowerbound2}.

\section{Randomized Algorithm for Common Lypunov Functions}
The randomized algorithm for the computation of $P$ in 
Assumption \ref{ass:subsystem_QA} 
is summarized here for the sake of completeness.

For a square matrix $X \in \mathbb{R}^{\sf n \times n}$, 
we denote its Frobenius norm by $\|X\|_F = 
(\sum_{i,k=1}^{\sf n} x_{i,k}^2)^{1/2}$, where $x_{i,k}$ is the $(i,k)$-th
entry of $X$.
For $X = X^{\top} \in \mathbb{R}^{\sf n \times n}$, 
let its eigenvalue decomposition be $X = U\Sigma U^{\top}$, where
$U$ is orthogonal and $\Sigma = \text{diag}(\lambda_1,\dots,\lambda_{\sf n})$.
For a fixed $\gamma \geq 0$, define $\Sigma_{\gamma} :=
\text{diag}(\max\{\lambda_1,\gamma\},\dots,\max\{\lambda_{\sf n},\gamma\})$
and set $G_{\delta,\delta_1}(X) := U\Sigma_{\gamma}U^{\top}$, where $\gamma :=
[(\delta^2 - \delta^2_1) / {\sf n}]^{1/2}$ for some $\delta > \delta_1 > 0$.

For the construction of 
{\em common} Lyapunov functions, we use
a scheduling function $h: \mathbb{Z}_+ \to \mathcal{P}$ that
has the following revisitation property~\cite{Liberzon2004}: 
For every element $i \in \mathcal{P}$ 
and for every integer $l \in \mathbb{Z}_+$, there exists an integer $k \geq l$ such that
$h(k) = i$. 

We can construct the common Lyapunov function in 
Assumption \ref{ass:subsystem_QA}
by using
the randomized algorithm of \cite{Ishii2004}, which
is based on
the gradient method proposed in \cite{Polyak2001}. 
\begin{alg}
	\label{alg:commonLyapunov}
		\begin{description}
			\item[(1)]
			Pick an initial $P^{[0,0]} > 0$ and set $R_0, r_0, \delta > 0$, and $\delta_1 \in (0,\delta)$.
			
			\item[(2)]
			%Find a finite subset of $\{\mathcal{Q}\}_{j \in \mathcal{S}}$ that covers
			%$\mathcal{B}(R_0)$ and
			%denote by $\mathcal{S}_N$ the index set of the cells in this subset.
			Find a finite index subset $\mathcal{S}_N$ of
			$\mathcal{S}$ such that 
			$\mathcal{B}(R_0) 
			\subset 
			\bigcup_{j \in \mathcal{S}_N}
			\mathcal{Q}_j$.
			
			\item[(3a)]
			Set $A := A_{h(k)}$, $B := B_{h(k)}$, and 
			$K := K_{h(k)}$, 
			and define
			\begin{align*}
				\phi(x_0,u,t) &:=
				e^{At}x_0 + \int^t_0 e^{A\tau} Bd\tau  \cdot u, \\
				u_j &:= Kq_j
				\\
				v(P,x,j,t) &:=
				(A\phi(x,u_j,t) + Bu_j)^{\top} P \phi(x,u_j,t) \\
				&\qquad + 
				\phi(x,u_j,t)^{\top} P (A\phi(x,u_j,t) + Bu_j) 
				 +
				C \| \phi(x,u_j,t) \|^2 \\
				\nabla_P v(P,x,j,t) &:=
				(A\phi(x,u_j,t) + Bu_j) \phi(x,u_j,t)^{\top} +
				\phi(x,u_j,t) (A\phi(x,u_j,t) + Bu_j)^{\top}\\
				\mathcal{X}_P(u) &:=
				\{
				x \in \mathbb{R}^{\sf n}:~
				(Ax+Bu)^{\top} Px  + x^{\top} P (Ax+Bu) \leq -C\|x\|^2
				\}.
			\end{align*} 
			
			\item[(3b)]
			Generate 
			\begin{align*}
				(x^{[k]},j^{[k]})
				&\in 
				\{
				(x,j): x \in [\mathcal{Q}_j \cap (\partial \mathcal{B}(r_0) \cup \partial \mathcal{B}(R_0))]  \cup (\partial \mathcal{Q}_j \cap \mathcal{B}(R_0)),~j \in \mathcal{S}_N
				\} \\
				 &=: \mathcal{F}
			\end{align*}
			according to some density function $f_{x,j}$ satisfying $f_{x,j}(x,j) > 0$ for all
			$(x,j) \in \mathcal{F}$.
			
			\item[(3c)]
			If $x^{[k]} \in \partial \mathcal{B}(r_0) \cup \partial \mathcal{B}(R_0)$, then
			set
			\begin{align*}
				&P^{[k+1,0]} = \\
				&\begin{cases}
					G_{\delta,\delta_1}(P^{[k,0]}) - \mu^{[k,0]} \nabla v^{[k,0]} & 
					\text{if $x^{[k]} \not\in \mathcal{X}_{P^{[k,0]}}(u_{j^{[k]}})$} \\
					P^{[k,0]} & \text{otherwise},
				\end{cases}
			\end{align*}
			where $\nabla v^{[k,0]} = \nabla_Pv(P^{[k,0]}, x^{[k]}, j^{[k]}, 0)$
			and $\mu^{[k,0]}$ is the step size given by
			\begin{equation*}
				\mu^{[k,0]} :=
				\frac{v(P^{[k,0]}, x^{[k]}, j^{[k]}, 0) + \delta \| \nabla v^{[k,0]}\|_F }{\|\nabla v^{[k,0]}\|_F^2}.
			\end{equation*}
			
			\item[(3d)]
			If $x^{[k]} \in \partial \mathcal{Q}_j \cap \mathcal{B}(R_0)$, then
			\begin{enumerate}
				\renewcommand{\labelenumi}{(\roman{enumi})}
				\item
				generate $\{ t^{[k,i]}\}^{l-1}_{i=0} \subset [0,T_s]$ according to
				some density function $f_t$ satisfying $f_t(t) > 0$ for all $t \in [0,T_s]$ with
				the indices in increasing order: $0 \leq t^{[k,0]} < \dots < t^{[k,l-1]} \leq T_s$;
				
				\item
				if $t^{[k,i]} \not=0$ and if $\phi(x^{[k]}, u_{j^{[k]}},t^{[k,i]} ) \in 
				\Cl(\mathcal{Q}_{j^{[k]}} \cap \mathcal{B}(R_0)^c \setminus \mathcal{B}(r_0) )$,
				then
				set $P^{[k+1,0]} = P^{[k,i]}$; otherwise set
				\begin{align*}
					&P^{[k,i+1]} \!=\! \begin{cases}
						G_{\delta,\delta_1}(P^{[k,i]}) \!-\! \mu^{[k,i]} \nabla v^{[k,i]} & \text{if $\phi(x^{[k]}, u_{j^{[k]}}, t^{[k,i]}) \!\not\in \!
							\mathcal{X}_{P^{[k,0]}}(u_{j^{[k]}}) \!\cup \! \mathcal{B}(R_0)$} \\
						P^{[k,i]} & \text{otherwise},
					\end{cases}
				\end{align*}
				where $\nabla v^{[k,i]} := \nabla_P v(P^{[k,0]}, x^{[k]}, j^{[k]}, t^{[k,i]})$
				is the step size given by
				\begin{equation*}
					\mu^{[k,i]} :=
					\frac{v(P^{[k,0]}, x^{[k]}, j^{[k]}, t^{[k,i]}) + \delta \| \nabla v^{[k,i]}\|_F }
					{\|\nabla v^{[k,i]}\|_F^2};
				\end{equation*}
				
				\item set $P^{[k+1,0]} = P^{[k,l]}$.
			\end{enumerate}
			
			\item[(4)]
			Find $R > 0$ satisfying 
			$\overline{\mathcal{E}}_{P^{[k,0]}}(R) \subset \mathcal{B}(R_0)$
			and obtain $r > 0$ satisfying $\mathcal{B}(r_0) \subset 
			\overline{\mathcal{E}}_{P^{[k,0]}}(r) \subset
			\overline{\mathcal{E}}_{P^{[k,0]}}(R)$ if it exists.
		\end{description}
\end{alg}

The major difference from  the algorithm in \cite{Ishii2004} 
is the procedure (3a), where
a scheduling function is used.
Under assumptions
similar to those in \cite{Ishii2004}, 
we can show that Algorithm \ref{alg:commonLyapunov}
gives a solution in a finite number of steps with probability one.
Since this is an immediate consequence of \cite{Ishii2004,Liberzon2004},
we omit the details.

\section*{Acknowledgment}  
The first author                             % Place acknowledgements
would like to thank Dr.
K. Okano of University California, Santa Barbara,
for helpful discussions.

%\appendix

\end{document}